\documentclass[a4paper,11pt]{article}
\usepackage{jcappub} 
\usepackage{lineno}
\usepackage{aas_macros}

\newcommand{\avg}[1]{\left\langle#1\right\rangle}
\DeclareUnicodeCharacter{2032}{$^\prime$}

\newif\iftrack
\iftrack
\newcommand{\added}[1]{{\bf #1}}
\newcommand{\deleted}[1]{}

\else
\newcommand{\added}[1]{{#1}}
\newcommand{\deleted}[1]{}

\fi

\arxivnumber{2309.03184}
\title{\boldmath Constraining the halo--ISM connection through multi-transition carbon monoxide line-intensity mapping}

\author{Dongwoo T.~Chung}
\affiliation{Canadian Institute for Theoretical Astrophysics, University of Toronto, 60 St.~George Street, Toronto, ON M5S 3H8, Canada}
\affiliation{Dunlap Institute for Astronomy and Astrophysics, University of Toronto, 50 St.~George Street, Toronto, ON M5S 3H4, Canada}

\emailAdd{dongwooc@cita.utoronto.ca}

\abstract{Line-intensity mapping (LIM) surveys will characterise the cosmological large-scale structure of emissivity in a range of atomic and molecular spectral lines, but existing literature rarely considers whether these surveys can recover excitation properties of the tracer gas species, such as the carbon monoxide (CO) molecule. Combining basic empirical and physical assumptions with the off-the-shelf Radex radiative transfer code or a Gaussian process emulator of Radex outputs, we devise a basic dark matter halo model for CO emission by tying bulk CO properties to halo properties, exposing physical variables governing CO excitation as free parameters. The CO Mapping Array Project (COMAP) is working towards a multi-band survey programme to observe both CO(1--0) and CO(2--1) at $z\sim7$. We show that this programme, as well as a further `Triple Deluxe' extension to higher frequencies covering CO(3--2), is fundamentally capable of successfully recovering the connection between halo mass and CO abundances, and constraining the molecular gas kinetic temperature and density within the star-forming interstellar medium in ways that single-transition CO LIM cannot. Given a fiducial thermal pressure of $\sim10^4$\,K\,cm$^{-3}$ for molecular gas in halos of $\sim10^{10}\,M_\odot$, simulated multi-band COMAP surveys successfully recover the thermal pressure within 68\% interval half-widths of 0.5--0.6 dex. Construction of multi-frequency LIM instrumentation to access multiple CO transitions is crucial in harnessing this capability, as part of a cosmic statistical probe of gas metallicity, dust chemistry, and other physical parameters in star-forming regions of the first galaxies and proto-galaxies out of reionisation.}

\begin{document}
\maketitle
\flushbottom
\section{Introduction}
\label{sec:intro}
Line-intensity mapping (LIM or IM) is a nascent mode of survey at the interface of extragalactic astrophysics and cosmology, targeting the unresolved aggregate spectral line emission of galaxies as opposed to individual emitters~\citep{LIM2017,LIM2019,BernalKovetz22}. Through the summary statistics of a spatial-spectral observation spanning cosmological scales, the data resulting from a LIM survey will distinguish three-dimensional modes of large-scale structure as illuminated by atomic or molecular line emission, tracing a particular species of gas and thus the elemental abundances and physical processes that excite the observed line emission.

The task is not without challenges, with interloper line emission and observational systematics all posing obstacles to achieving background-limited sensitivity. However, all LIM surveys---and indeed all cosmological surveys---trace the same underlying large-scale structure lit up in different but correlated ways. As such, cross-correlations have led to the first detections of cosmological neutral atomic hydrogen emitting in the 21 cm line~\citep{Chang10,Masui13} as well as tentative hints of ionised carbon line emission~\citep{Pullen18,Yang19}, by leveraging external cross-correlations against galaxy surveys. LIM experiments targeting other gas species like carbon monoxide (CO) have also analysed existing data for possible external cross-correlations~\citep{Keenan22,COMAPESVIII}, and expect to achieve sufficient sensitivity for a strong ($\gtrsim5$ sigma) cross-correlation detection well ahead of any auto-correlation detection of equally high significance~\citep{COMAPESV_}. (Interferometric CO LIM surveys have also claimed auto-correlation detections at the 2--3 sigma level~\citep{Keating16,mmIME-ACA}.)

Importantly, a cross-correlation result is more than the sum of two auto-correlation results. Beyond the practical advantage of rejecting disjoint foregrounds and systematics---thus removing obstacles to an initial demonstration of detectability of cosmological line emission---cross-correlation between two data sets enables science not possible with either data set in isolation. Cross-correlation between LIM and galaxy surveys, for instance, will enable insights into the atomic and molecular gas content of the cross-correlated galaxies~\citep{Wolz17,selfcite_COMAPHX,BreysseAlexandroff19,Silva21}. Just as important are LIM--LIM cross-correlations, particularly between 21\,cm IM and cm- to mm-wave LIM in star-formation lines~\citep{Beane19,Padmanabhan23,McBrideLiu23}, but also between different star-formation lines~\citep{BreysseRahman17,LiuBreysse21,Sun22,Padmanabhan22}---in some cases even within the same survey. For instance, the CO Mapping Array Project (COMAP;~\cite{COMAPESI}), a single-dish CO LIM experiment, is currently in a Pathfinder phase operating at observing frequencies of 26--34 GHz and primarily targeting CO(1--0) at $z\sim3$. However, as both $^{12}$CO and $^{13}$CO emission fall within these observing frequencies, an internal cross-correlation across the appropriate frequency channels could enable insights into isotopologue ratios and saturated fractions~\citep{BreysseRahman17}.

\begin{figure}
    \centering
    \includegraphics[width=0.86\linewidth]{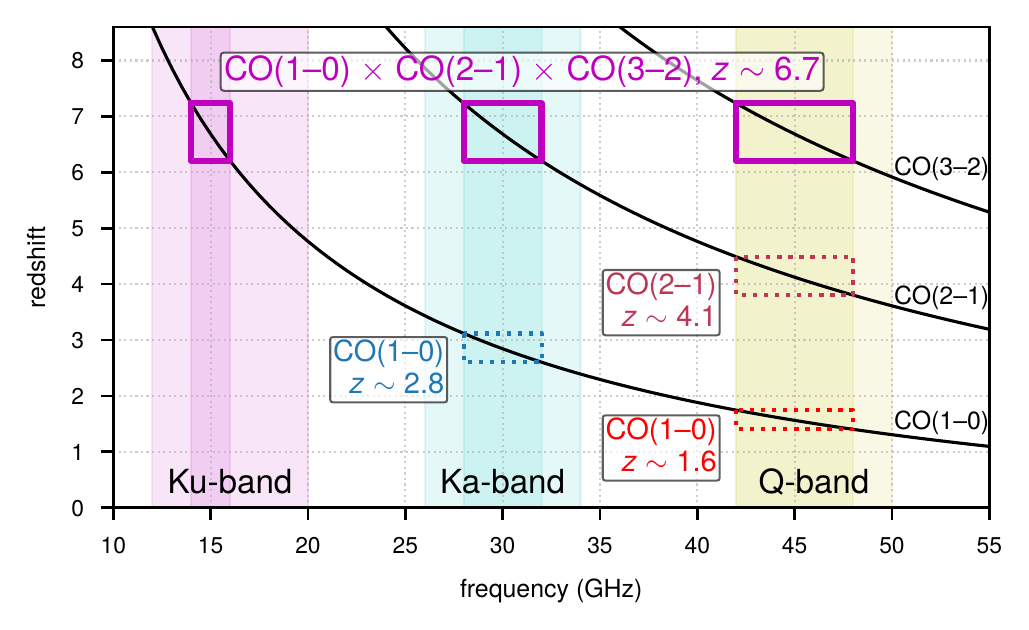}
    \caption{Illustration of the correspondence of observing frequency with the emission redshift of the three lowest-energy CO rotational transition lines. We also show the possible observational bands for future COMAP phases considered in simulated scenarios in this work, as well as the specific target and interloper components (solid and dotted boxes respectively) for reionisation-epoch CO LIM observations.}
    \label{fig:ladder}
\end{figure}

The benefits of cross-correlation motivate the fiducial design of future COMAP iterations beyond the current Pathfinder phase~\citep{Breysse22}, involving an expansion not only of existing instrumentation centred at 30 GHz, but also construction of new spectrometers observing at 15 GHz. As~\autoref{fig:ladder} shows, this enables extraction of CO emission at $z\sim7$---towards the end of the epoch of reionisation (EoR)---through cross-correlation of the lowest rotational energy transitions of the CO molecule. Any detection will be robust not only to excitation conditions at the emission epoch that may suppress higher energy transitions, but also to disjoint foregrounds. This includes CO(1--0) emission at $z\sim3$---the primary target for the Pathfinder, but a prominent foreground in the way of surveying CO(2--1) emission at $z\sim7$.

However, the benefits of multi-transition CO LIM should extend beyond merely improving robustness of a late-reionisation CO detection. Much in the same way that line ratios derived from stacking resolved sources sample the spectral line energy distribution (SLED) of the galaxies and thus reveal the average properties of the interstellar medium (ISM) of the population (e.g.,~\cite{Spilker14,Boogaard20,Reuter23}), multi-transition LIM data will effectively sample the SLED of all contributing unresolved sources. By probing multiple transitions of CO, the data from future iterations of COMAP will have constraining power over the physical conditions in the ISM of late-reionisation galaxies that govern the excitation of these CO rotational transitions, such as the density and kinetic temperature of the surrounding medium as populated by molecular hydrogen (H$_2$), the primary collisional partner for CO.

Accounting for such conditions is key to a complete understanding of the physics of star formation towards the end of reionisation. To successfully recover CO and H$_2$ densities at these epochs, even in the `cosmic average' molecular cloud, is to understand the conversion between the two and thus the metallicity and turbulence in these environments (cf.~the review of the CO-to-H$_2$ conversion factor by~\cite{Bolatto13}), with turbulence in particular governing cloud collapse and fragmentation. Constraints on parameters like gas kinetic temperature will be especially powerful in combination with measurements of the redshift evolution of dust temperature (e.g.,~\cite{Viero22}). As feedback processes influence the complex chemical interplay of molecular gas and dust in dense environments (cf., e.g.,~\cite{Hocuk16,Hu23}), CO measurements at these high redshifts will provide important insights into cooling and heating mechanisms governing star formation at the very inception of the star-forming cosmic web.

The addition of instrumentation outside the 26--34 GHz COMAP Pathfinder frequency range is thus transformative, not incremental. The future may see not only a `double deluxe' COMAP-EoR survey and a COMAP Extended Reionisation Array (COMAP-ERA) incorporating 15 GHz and 30 GHz observations (in the so-called Ku and Ka bands)---as outlined by the forecasts of~\cite{Breysse22}---but in fact a COMAP `Triple Deluxe' also targeting CO(3--2) at the same late-reionisation redshifts, with additional 45 GHz instrumentation (in the so-called Q band). Taking full advantage of this transformative capability will however require consistent modelling of line ratios across the CO SLED grounded in at least a statistical picture of the physical conditions of the ISM.

Yet previous literature forecasting CO LIM signals at $z\gtrsim6$~\citep{Lidz11,Gong11,MDK19,Yang22,COMAPESVIII} leans towards either modelling only a single CO transition, or applying empirical halo models for multiple CO transitions that ultimately abstract away many of the physical variables governing CO excitation in dense gas. That said, some works have constructed modelling frameworks for LIM observables encompassing multiple lines based on explicit, self-consistent physical approaches (e.g.,~\cite{Mashian15,Sun19,SIDES22,LIMFAST_I,LIMFAST_II}). For quite possibly the first time, we move beyond even those works to present a detailed consideration of recoverability of ISM environmental variables specifically from multi-transition CO LIM.

To quantify the potential impact of multi-transition CO LIM in idealised scenarios, we consider here explicit calculations of observables and observabilities for different iterations of COMAP targeting late reionisation, rooted in self-consistent modelling of the CO SLED aided by the off-the-shelf non-LTE radiative transfer code Radex~\cite{RADEX} (specifically the implementation of~\cite{spectralradex}). In doing so, we aim to answer the following questions:
\begin{itemize}
    \item What constraining power can future CO line-intensity mapping provide on the environmental variables that govern CO excitation?
    \item How much does this constraining power depend on the ability of future surveys to observe multiple lines, in particular multiple transitions of CO?
    \item Do qualitative aspects of answers to the above change with our approach to modelling the high-redshift CO signal?
\end{itemize}

The paper is structured as follows. After detailing methods for simulations of signal, noise, and parameter recovery in~\autoref{sec:methods}, we present detectability and parameter recovery forecasts in~\autoref{sec:results}. We discuss key takeaways in~\autoref{sec:discussion} before concluding in~\autoref{sec:conclusions}.

Unless otherwise stated, we assume base-10 logarithms, and a $\Lambda$CDM cosmology with parameters $\Omega_m = 0.307$, $\Omega_\Lambda = 0.693$, $\Omega_b =0.0486$, $H_0=100h$\,km\,s$^{-1}$\,Mpc$^{-1}$ with $h=0.677$, $\sigma_8 =0.816$, and $n_s =0.967$, consistent with the findings of the Planck satellite (based on~\cite{Planck15} but still consistent even with~\cite{Planck18}). Distances carry an implicit $h^{-1}$ dependence throughout, which propagates through masses (all based on virial halo masses, proportional to $h^{-1}$) and volume densities ($\propto h^3$).

\section{Model for simulated observations}
\label{sec:methods}
\subsection{Signal: mocking the halo--CO connection}
We consider two possible ways to model the signal, i.e., to model CO line fluxes per dark matter halo. One is an entirely empirical model of the bulk CO column density and other properties as a function of halo mass and redshift, while the other attempts to provide some physical motivation by modelling a population of molecular clouds and formulating a per-cloud prescription of ISM properties. We will dub these Model A (for `amalgamated') and Model B (for `broken-down') respectively.

Across both models, we assume a background temperature of $(1+z)\times2.7255$ K with $z=6.68$ set to the redshift observed at 15 GHz in CO(1--0) or 30 GHz in CO(2--1). However, the Radex calculation of CO line fluxes from a homogeneous object demands definition of further properties, namely a geometry for the escape probability, the kinetic temperature $T_k$, H$_2$ volume density $n_\text{H2}$, the line width $\Delta v$, and the CO column density $N_\text{CO}$. We assume throughout that the kinetic temperature in environments hosting CO is the same, at 86 K. The choice is motivated by the recent analysis of~\cite{Viero22}; at $z\sim7$ their model suggests a dust temperature of $\approx86$ K and we expect the kinetic temperature to be at least weakly coupled to this temperature. The choices of all other parameters will differ significantly between Models A and B owing to their divergent approaches to describing the CO content of dark matter halos.

\subsubsection{Model A: amalgamated CO properties per dark matter halo}
The key assumption at the core of Model A is the assumption that the CO column density follows a double power law with halo mass:
\begin{equation}N_\text{CO}(M_h)=\frac{N_\text{break}}{(M_h/M_\text{break})^{2/3}+(M_h/M_\text{break})^{-\alpha}}.\end{equation}
At low masses this assumes a power law of index $\alpha$, and at high masses the power law assumes a fixed index of $-2/3$. The resulting CO line fluxes per halo are then proportional to this column density scaled by various physical variables, including population fractions at each energy level.

The motivation for choosing the high-mass power law index of $-2/3$ is to effect a saturation of area-integrated CO luminosity per halo at high halo mass. Following \cite{Mashian15}, we assume that the CO emission and star formation both originate from a disc with radius $R_d=(\lambda/\sqrt{2})(j_d/m_d)r_\text{vir}$, with $\lambda=0.05$ (isolated exponential profile) and $j_d/m_d=1$ (same specific angular momentum for disc and halo). Since the total CO luminosity is the product of the velocity-integrated CO flux with the area of this disc, it thus scales as $N_\text{CO}(M_h)\times M_h^{2/3}$, and should then plateau at high $M_h$ given our form of $N_\text{CO}(M_h)$.

We assume that $M_\text{break}$ lies where the star formation rate peaks, around $10^{12.3}\,M_\odot$. We also derive a fiducial value of $\alpha=0.9$ based on the following sequence of considerations:
\begin{itemize}
    \item The best-fit UniverseMachine model from~\cite{UM} finds that the SFR at low masses evolves as $v_\text{peak}^5$ at COMAP-EoR redshifts. If $v_\text{peak}\sim M_h^{1/3}$, then the SFR evolves as $M_h^{5/3}$.
    \item The SFR surface density then must scale roughly linearly with $M_h$, since the area over which this SFR occurs scales as $R_d^2\sim M_h^{2/3}$.
    \item If the CO surface density follows a Kennicutt--Schmidt law with $\Sigma_\text{SFR}\sim N_\text{CO}^n$ with $n$ somewhere between 1 and 1.2~\cite{Genzel10,Leroy13}, then at the faint end this yields $N_\text{CO}\sim M_h^{1/n}$, so that $\alpha=1/n$ is somewhere between 1 and 0.8. The choice of $\alpha=0.9$ is the midpoint of this range.
\end{itemize}
We also assume a minimum CO emitter halo mass of $M_\text{min}=10^{10}\,M_\odot$, consistent with the photoionisation considerations in~\cite{Mashian15} introducing a sharp cutoff at this halo mass.

Radex provides three options for escape probability geometry: a uniform sphere, a turbulent sphere (equivalent to a large velocity gradient approximation), and a plane-parallel slab. We use the first option for Model A.

We also assume a rather strong evolution of the H$_2$ volume density with the host halo mass:
\begin{equation}n_\text{H2}(M_h) = n_\text{H2,10}\left(\frac{M_h}{10^{10}\,M_\odot}\right)^{2\alpha},\end{equation}
where $\alpha$ is the same $\alpha$ found in the definition of $N_\text{CO}(M_h)$. The form is motivated by the expectation discussed by~\cite{Mashian15} that the volume density will go as the square of the surface density, whereas the column density scales linearly with surface density. We do not prescribe a break here because molecular hydrogen is more numerous and thus more capable of self-shielding than CO, and expect this to be the case for the mass range that dominates the signal. Our fiducial value for $n_\text{H2,10}$ is $10^2$\,cm$^{-3}$.

With all of these inputs, Radex can calculate the velocity-integrated flux $F_J$ per halo in each CO($J\to J-1$) line, leading to the luminosity per halo being $L'_J = \pi R_d^2F_J$.

\subsubsection{Model B: break-down of the CO content of halos into molecular clouds}
This model attempts to model the CO content of each dark matter halo as the ensemble of some number of molecular clouds, which altogether make up the molecular mass of the galaxy. For this, we take additional inspiration from the work of~\cite{Popping19}, which couples the Santa Cruz semi-analytic model of galaxy formation to a radiative transfer code to model a range of carbonic lines.

For simplicity we assume all clouds are near-virialised, and furthermore have the same mass and size, taking the properties of the average cloud from the cloud mass function assumed by~\cite{Popping19}. The power-law (truncated Pareto) form of that cloud mass function yields an average cold gas mass per cloud of $M_\text{MC}\approx2.2\times10^4\,M_\odot$. Based on the virial theorem, there is an expectation that the external pressure is related to the mass and radius of the molecular cloud, and in particular $P_\text{ext}\sim M_\text{MC}^2r_\text{MC}^{-4}$ (see, e.g.,~\cite{Elmegreen89}). Taking the external pressure acting on the cloud to be $P_\text{ext}/k_B\sim10^4\,$K\,cm$^{-3}$,\footnote{It is conventional to express pressure in interstellar environments in units of temperature times number density; for reference, $P_\text{ext}/k_B\sim10^4$\,cm$^{-3}$\,K is equivalent to $P_\text{ext}\sim10^{-15}\,\text{torr}\approx140$\,fPa.} the cloud size for this mass is of order $r_\text{MC}\approx8.7$ pc based on the calibration given by~\cite{Popping19} (in turn taken from~\cite{Swinbank11}). This in turn naturally suggests $n_\text{H2}\approx110$\,cm$^{-3}$ (see~\autoref{eq:nH2})---so now changing this parameter equates to changing the cloud size by modulating the environmental pressure necessary to arrive at the given density.

Incidentally, the corresponding molecular hydrogen column density is $N_\text{H2}\sim3\times10^{21}$ cm$^{-2}$, or a mass surface density of $\Sigma\sim90\,M_\odot\,$pc$^{-2}$. As a point of comparison, the work of~\cite{Miville17} catalogued molecular clouds across the entire Milky Way disk and found their surface densities to span a range of 2 to 300\,$M_\odot$\,pc$^{-2}$.

\cite{Miville17} also obtained a fitting formula for the velocity dispersion $\sigma_v$ in terms of $\Sigma$ and $r_\text{MC}$:
\begin{equation}
    \frac{\sigma_v}{\text{km\,s}^{-1}} = 0.23\left(\frac{r_\text{MC}\Sigma}{M_\odot\,\text{pc}^{-1}}\right)^{0.43}
\end{equation}
This suggests $\sigma_v\sim4$\,km\,s$^{-1}$ for our average cloud. We assume a Gaussian line velocity profile with this $\sigma_v$, leading to a full width at half maximum (FWHM) of $\Delta v=10$ km s$^{-1}$. This value is similar to the effective sound speed assumed by~\cite{Mashian15}. As~\cite{Miville17} note, the origin of the line width and the nature of its relation to other cloud parameters is the subject of some debate~\citep{BP11}, so we avoid coupling $\Delta v$ too tightly to our assumption of virial clouds, and leave it fixed at the fiducial value of $10$\,km\,s$^{-1}$ throughout.\footnote{Ultimately, the relevant free parameter in this model would be $n_\text{H2}$, and as we will see $N_\text{H2}r_\text{MC}\sim n_\text{H2}^{1/3}$, implying $\sigma_v\sim n_\text{H2}^{0.14}$. This is an extremely weak dependence compared to the level of scatter observed by~\cite{Miville17} in the correlation between line width, radius, and surface density.}

The simulation work of~\cite{Popping19} suggests that the evolution of H$_2$ mass with increasing halo mass is roughly linear up to $M_h\sim10^{12}\,M_\odot$, then plateaus. We also take cues from the work of~\cite{NeutralUM}, starting with the \textsc{UniverseMachine} empirical approach of modelling star-formation histories and applying it to neutral atomic and molecular gas histories. That work suggests a slightly superlinear growth of molecular gas mass with halo mass at high redshift: the ratio of H$_2$ mass to halo mass grows roughly as $M_h^{0.5}$, although again shrinking slightly at higher masses. We thus consider the cloud count to saturate at high halo mass, and take
\begin{equation}N_\text{MC}(M_h) = \frac{1}{2.2\times10^4\,M_\odot}\times\underbrace{\frac{4.2\times10^{11}\,M_\odot}{1+(M_h/M_\text{brk})^{-3/2}}}_{M_\text{H2}(M_h)}\end{equation}
with $\log(M_\text{brk}/M_\odot)=12.3$ as before. The denominator implying a H$_2$ mass of $2.1\times10^{11}\,M_\odot$ for a halo of mass $M_\text{brk}\approx2\times10^{12}\,M_\odot$ is tuned to yield a cosmic H$_2$ mass density of $2.5\times10^7\,M_\odot$ Mpc$^{-3}$ at $z=6.7$.

For comparison, the only observational constraint that approaches these redshifts comes from the CO Luminosity Density at High Redshift (COLD$z$) survey~\citep{COLDz,COLDzLF}, an interferometric untargeted search for CO line emitter candidates in CO(1--0) at $z\sim2$--3 and CO(2--1) at $z\sim5$--7. The most conservative upper limit, assuming all unconfirmed emission identified as $z\sim3$ CO(1--0) lines could be $z\sim7$ CO(2--1) lines, is $\rho_\text{H2}<4.0\times10^7\,M_\odot$ Mpc$^{-3}$. Our fiducial functions for cloud mass and count per halo thus result in a value consistent with the primary observational constraint at this redshift, which also depends strongly on extrapolation (if any) of the CO luminosity function beyond the survey detection limit (noting that the COLD$z$ analysis of \cite{COLDzLF} carries out no such extrapolation), and on assumptions around CO excitation and the CO-to-H$_2$ conversion factor.

The halo mass modulates the CO column density per cloud through its connection to the hosted gas-phase metallicity. Like~\cite{Mashian15}, we assume that all carbon in molecular clouds is locked up in CO on average, taking the most optimistic end of the 30--100\% range considered by the work of~\cite{BialySternberg15}, which is also how~\cite{Mashian15} obtains the conversion from H$_2$ column density to CO column density as
\begin{equation}
    \chi_\text{CO} = 3\times10^{-4}(Z'/Z_\odot).
\end{equation}
We assume $Z'\simeq1$ (in units of solar metallicity) at $M_h\simeq M_\text{brk}$, with the evolution essentially following a double power law with the same break scale as the cloud count:
\begin{equation}
    \frac{Z'(M_h)}{Z_\odot} = \frac{Z'_\text{brk}}{1+(M_h/M_\text{brk})^{-\alpha}}\end{equation}
Our fiducial assumption is $\alpha=1/3$ and $Z'_\text{brk}=2$ so that $Z'(10^{10}\,M_\odot)\approx0.25$ and $Z'(10^{9}\,M_\odot)\approx0.125$. In general we assume that $\alpha>0$.

Because the metallicity gives $\chi_\text{CO}$, combining this with $n_\text{H2}$ forces a value for $N_\text{CO}$. To be explicit:
\begin{equation}\frac{4}{3}\pi r_\text{MC}^3n_\text{H2}\simeq\pi r_\text{MC}^2N_\text{H2},\end{equation}
meaning
\begin{equation}N_\text{CO} = \chi_\text{CO}N_\text{H2} = 4\times10^{-4}\times n_\text{H2}r_\text{MC}Z'(M_h)/Z_\odot.\end{equation}
But furthermore, we derive $r_\text{MC}$ anyway from the number density, size, and total mass of each molecular cloud:
\begin{equation}n_\text{H2}=\frac{2.2\times10^4\,M_\odot}{\mu m_\text{H2}(4/3)\pi r_\text{MC}^3};\label{eq:nH2}\end{equation}
here $\mu m_\text{H2}$ is the mass per molecule of molecular gas, where following~\cite{Mashian15} we multiply the H$_2$ molecular mass by a factor of $\mu=1.36$ to account for the occasional contribution of helium to the total mass of cold gas. We invert this to obtain
\begin{equation}r_\text{MC}=\left(\frac{2.2\times10^4\,M_\odot}{\mu m_\text{H2}(4/3)\pi n_\text{H2}}\right)^{1/3}=1.3\times10^{20}\times n_\text{H2}^{-1/3}.\end{equation}
As a result, our expression for the CO column density simplifies into
\begin{equation}N_\text{CO} = \chi_\text{CO}N_\text{H2} = 5.2\times10^{16}\times n_\text{H2}^{2/3}Z'(M_h)/Z_\odot.\label{eq:NCO_modelB}\end{equation}
For our fiducial model parameter values, this ends up in the range of $\sim10^{17}$--$10^{18}$ cm$^{-2}$ for typical halo masses of $10^{9}$--$10^{12}\,M_\odot$.

Note that several of our model parameters end up in a degenerate combination as an overall normalisation of column density at this step, namely
\begin{itemize}
\item the H$_2$ mass $M_\text{MC}=2.2\times10^4\,M_\odot$ per cloud;
\item the metallicity at break mass, $Z'_\text{brk}/2=1$;
\item the CO--H$_2$ conversion factor at solar metallicity, $\chi_\text{CO,1}=3\times10^{-4}$;
\item and the H$_2$ volume density $n_\text{H2}=110$ cm$^{-3}$ in each cloud.
\end{itemize}
We merge the combination of the first three into a single parameter:
\begin{equation}
    N_\text{CO,brk}\equiv\frac{Z'_\text{brk}}{2}\times\frac{4\chi_\text{CO,1}}{3}\times\left(\frac{M_\text{MC}}{\mu m_\text{H2}(4/3)\pi}\right)^{1/3}
\end{equation}

All in all, the model follows these steps:
\begin{itemize}
\item Assume that only halos above some minimum mass $M_\text{min}=10^{9}\,M_\odot$ host CO emission. (Note that this is a different value from that assumed for Model A.)
\item Assume the escape probability geometry of a turbulent sphere.
\item Assume the same kinetic temperature for each cloud, with the fiducial value again being $T_k=86$ K. Note that this is completely decoupled from the virial temperature of the molecular cloud, as we do not assume that the thermal energy (as opposed to magnetic or turbulent kinetic energy) is what primarily counteracts the gravitational potential energy or any external confining pressure.
\item Assume the same $n_\text{H2}$ for each cloud, with the fiducial value being 110 cm$^{-3}$.
\item The CO column density per cloud as a function of host halo mass is ultimately
\begin{equation}N_\text{CO}(M_h) = \frac{2N_\text{CO,brk}n_\text{H2}^{2/3}}{1+(M_h/M_\text{brk})^{-\alpha}}.\end{equation}
\item The number of clouds at each halo mass is given by
\begin{equation}N_\text{MC}(M_h)=\frac{2N_\text{MC,brk}}{1+(M_h/M_\text{brk})^{-\beta}},\end{equation}
where the fiducial values of the free parameters are $M_\text{brk}=10^{12.3}\,M_\odot$, $N_\text{MC,brk}=9.5\times10^6$, and $\beta=3/2$, based on the assumptions made above about the ratio of H$_2$ mass to halo mass and the H$_2$ mass per cloud.
\end{itemize}
The CO($J\to J-1$) luminosity per cloud, phrased as the velocity- and area-integrated brightness temperature, then becomes $\pi r_\text{MC}^2$ times the velocity-integrated flux $F_{J,\text{MC}}$ obtained from Radex. Although $r_\text{MC}$ depends on both the mass per cloud and the number density, here we will only account for the latter dependence, effectively fixing $M_\text{MC}$. In this case, $r_\text{MC}\approx 42$\,pc when $n_\text{H2}=1$\,cm$^{-3}$. We also assume that the luminosities of each cloud add together linearly to result in the total luminosity per halo, meaning that for each line
\begin{equation}
    L'_J(M_h)=N_\text{MC}(M_h)\times\pi \left(\frac{42\,\text{pc}}{n_\text{H2}^{1/3}}\right)^2F_{J,\text{MC}}.
\end{equation}
Seven free parameters $\{T_k,n_\text{H2},M_\text{brk},N_\text{MC,brk},N_\text{CO,brk},\alpha,\beta\}$ control this relation in the end, with fiducial values of $\{86\,$K$,110\,$cm$^{-3},10^{12.3}\,M_\odot,9.5\times10^6,5.2\times10^{16},1/3,3/2\}$.

\begin{figure}
    \centering
    \includegraphics[width=0.5\linewidth]{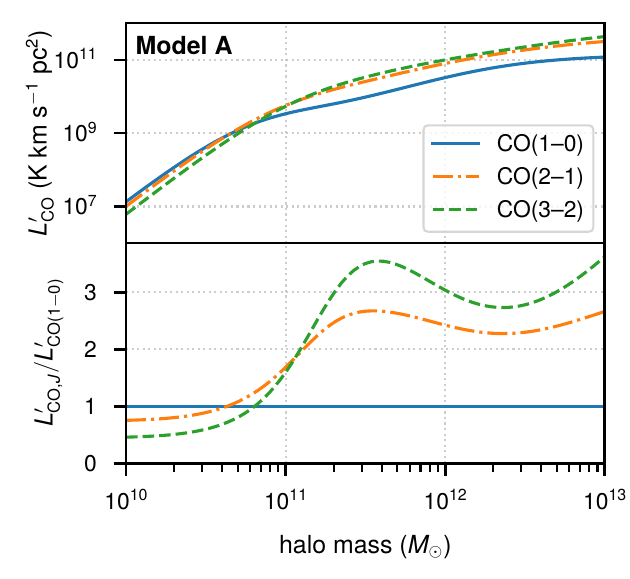}\includegraphics[width=0.5\linewidth]{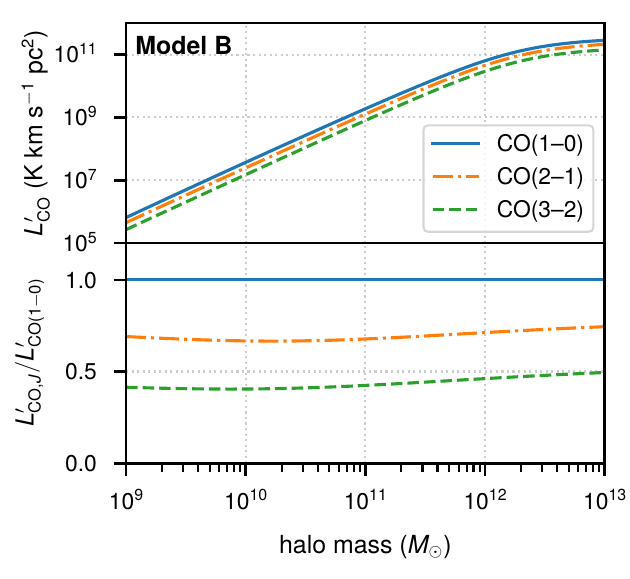}
    \caption{Summaries of Model A (left) and Model B (right), showing the relations between halo mass and line luminosity (upper panels) and ratios between line luminosities (lower panels) given fiducial parameter values.}
    \label{fig:LofM}
\end{figure}

Before we move on to how these models of $L'_\text{CO}(M_h)$ lead to observables like the power spectra of the CO lines, we provide a few summaries of Model A and Model B. We summarise the parameters and their fiducial values in~\autoref{tab:modelsummary}, and the resulting $L'_J$ relations and ratios between $L'_J$ and $L'_1$ in~\autoref{fig:LofM}. From the latter, in particular, we note that Model A predicts that the excitation of CO evolves from sub-thermal at low halo masses to super-thermal at high masses, owing to the strong evolution of $n_\text{H2}(M_h)\sim M_h^{2\alpha}=M_h^{1.8}$ with halo mass. Model B on the other hand, with $n_\text{H2}$ kept the same across all clouds in all halos, essentially predicts sub-thermal excitation across the entire halo mass range with little evolution.

\added{We also show in~\autoref{sec:modelvar} the effect of parameter variations in Model B, on both the $L_J'(M_h)$ relation shown in~\autoref{fig:LofM} and on the power spectra calculated in the next section.}

\begin{table}
    \centering
    \begin{tabular}{c|c|c|c}
        Model & Variable & Fiducial value & Meaning\\\hline
        A & $\alpha$ & 0.9 & Low-mass power-law index for $N_\text{CO}(M_h)$ relation\\
        &&& (half of power-law index for $n_\text{H2}(M_h)$ relation)\\
        & $M_\text{break}$ & $10^{12.3}\,M_\odot$ & Characteristic mass scale for $N_\text{CO}(M_h)$ relation\\
        & $N_\text{break}$ & $10^{19.86}$\,cm$^{-2}$& Total CO column density in halo with $M_h=M_\text{brk}$\\
        & $T_k$ & 86 K & Kinetic temperature \\
        & $n_\text{H2,10}$ & $10^2$\,cm$^{-3}$ & H$_2$ volume density in halo of mass $10^{10}\,M_\odot$\\\hline
        B & $\alpha$ & 1/3 & Low-mass power-law index for $N_\text{CO}(M_h)$ relation\\
        & $M_\text{brk}$ & $10^{12.3}\,M_\odot$ & Characteristic mass scale for $N_\text{CO}(M_h)$ relation\\
        & $N_\text{CO,brk}$ & $10^{16.72}$& Normalisation for CO column density per cloud\\
        & $T_k$ & 86 K & Kinetic temperature \\
        & $n_\text{H2}$ & $110$\,cm$^{-3}$ & H$_2$ volume density in each molecular cloud\\
         & $\beta$ & 3/2 & Low-mass power-law index for $N_\text{MC}(M_h)$ relation\\
        & $N_\text{MC,brk}$ & $9.5\times10^6$ & Number of molecular clouds in halo of mass $M_\text{brk}$\\
        & $M_\text{min}$ & $10^{9}\,M_\odot$ & Minimum halo mass for CO emission\\
        & & & (fixed at $10^{10}\,M_\odot$ for Model A)
    \end{tabular}
    \caption{Parameters allowed to vary for both Models A and B throughout this work. Some parameters relate to the CO column density $N_\text{CO}(M_h)$ (per halo for Model A, per molecular cloud for Model B) or the number of molecular clouds per halo $N_\text{MC}(M_h)$ (for Model B) as functions of halo mass.}
    \label{tab:modelsummary}
\end{table}
\subsubsection{Observables: auto and cross power spectra}
We consider auto and cross power spectra across the different CO transitions. We specifically only consider the spherically averaged 3D power spectrum $P(k)$ as a function of comoving wavenumber $k$; while~\cite{Breysse22} also considered the quadrupole of the 3D power spectrum, we conservatively model only the `monopole' $P(k)$, assuming that it is the easiest to detect and correlate across all bands. Otherwise we draw heavily from the formalism of~\cite{Breysse22} below.

We translate the CO luminosities $L'_J(M_h)$ as functions of halo mass into the observable power spectra using the fit of~\cite{Tinker08} for the halo mass function $dn/dM_h$ (the differential comoving number density of halos of mass $M_h$) and the model of~\cite{Tinker10} for the halo bias $b(M_h)$ with which halos of mass $M_h$ trace the matter density contrast. Given the CO($J\to J-1$) luminosity $L_J'(M_h)$ as a function of halo mass, we can find the average line brightness temperature,
\begin{equation}
    \avg{T}_J=\frac{(1+z)^2}{H(z)}\int dM_h\,\frac{dn}{dM_h}\,L'_J(M_h),\label{eq:Tavg}
\end{equation}
and the average temperature--bias product,
\begin{equation}
    \avg{Tb}_J=\frac{(1+z)^2}{H(z)}\int dM_h\,\frac{dn}{dM_h}\,L'_J(M_h)b(M_h),
\end{equation}
with $H(z)$ being the redshift-dependent Hubble parameter for our fiducial cosmology. The luminosity-weighted average bias of the CO emission is then $b_J=\avg{Tb}_J/{\avg{T}_J}$.

In real comoving space, the spherically averaged auto power spectrum for any transition is the sum of a clustering component proportional to the matter power spectrum $P_m(k)$ and a scale-independent shot noise component $P_s$:
\begin{equation}
    P_{J\times J}(k) = \avg{Tb}_JP_m(k)+P_{s,J},
\end{equation}
where
\begin{equation}
    P_{s,J} = \frac{(1+z)^4}{H^2(z)}\int dM_h\,\frac{dn}{dM_h}\,L^{\prime2}_J(M_h).\label{eq:Pshot_auto}
\end{equation}
(Note that we do not assume any halo-to-halo scatter around the mean $L'_J(M_h)$ relation\added{;~\autoref{sec:modelvar} discusses the effect of adding such scatter to model predictions}.)

The actual observed auto power spectrum undergoes a number of anisotropic distortions due to large-scale structure flows, line broadening, and angular resolution. Accounting for all such distortions results in a 2D auto pseudo-power spectrum $\tilde{P}(k,\mu)$, dependent on both $k$ and $\mu$ (the cosine of the angle of the wavevector relative to the line-of-sight direction, such that the line-of-sight component is $k_\parallel = k\mu$ and the transverse component is $k_\perp=k\sqrt{1-\mu^2}$). It is necessary to then average over $\mu\in(-1,1)$ to obtain the spherically averaged pseudo-power spectrum $\tilde{P}(k)$ for each line.

Assuming both the beam profile and line profile are Gaussian functions, the attenuation due to these two factors is given by
\begin{equation}
    W_r(k,\mu) = \exp{\left[-k^2\sigma_\perp^2(1-\mu^2)-k^2\sigma_\parallel^2\mu^2\right]}.
\end{equation}
The transverse profile is defined by $\sigma_\perp=\chi(z)\theta_B/\sqrt{8\ln{2}}$, where the comoving angular diameter distance to redshift $z$ is equal to simply the comoving distance $\chi(z)$ to redshift $z$ for our fiducial flat cosmology, and $\theta_B$ is the angular FWHM of the main beam. We assume values of 3.7/3.9/4.0 arcminutes for $\theta_B$ at 15/30/45 GHz, which are possibly slightly optimistic at the level of 10--20\%, but nonetheless realistically achievable with dishes of diameters in the range of 24--7 m. For reference, the dishes expected to be used for COMAP-EoR are 18\,m class at 15 GHz and 10\,m class at 30 GHz.

Meanwhile, the line-of-sight profile is given by
\begin{equation}
    \sigma_\parallel=\frac{(1+z)}{H(z)}\frac{v_\text{eff}}{\sqrt{8\ln{2}}},\label{eq:vfwhmtosig}
\end{equation}
where here we assume a typical line width of $v_\text{eff}=250$ km s$^{-1}$. Note that this represents the velocity dispersion per halo, not per cloud, and therefore we apply it equally to Models A and B.

We also put in two additional factors to account for loss in sensitivity due to survey parameters. The first is the main beam efficiency $\eta$ of the instrument, which will vary by band depending on instrumental details like blockage due to on-axis reflectors. We follow the values of $\eta_1=1$ and $\eta_2=0.72$ for the 15 GHz and 30 GHz instrumentation assumed in~\cite{Breysse22}, and pick a value of $\eta_3=0.86$ roughly in the middle for the 45 GHz instrumentation (assuming effectively that moderate improvements can be made over the 30 GHz instrumentation even in the case of on-axis antennas being used). The loss of signal manifests as a factor of $\eta_J$ in temperature and thus a factor of $\eta_J^2$ in the auto power spectrum.

The second is a volume window function, following~\cite{Bernal19}, to account for the loss of modes at scales approaching the limits of the survey volume. Here we assume that the full overlap between all data sets across the three transitions spans $z=6.2$--7.2, corresponding to observations of CO(1--0) at 14--16 GHz, and fields of size $\Omega_F=4$ deg$^2$. Assuming then that the maximum accessible transverse scale is $L_\perp=\chi(z=6.7)\sqrt{\Omega_F}$ (the transverse span of each survey patch at the central redshift) and the maximum accessible line-of-sight scale is $L_\parallel=\chi(z=7.2)-\chi(z=6.2)$, the corresponding minimum accessible wavenumbers are $k_{\perp,\text{min}}=2\pi/L_\perp$ and $k_{\parallel,\text{min}}=2\pi/L_\parallel$, and the volume window function is
\begin{equation}
    W_V(k,\mu)= \left\{1-\exp{\left[-\left(\frac{k}{k_{\perp,\text{min}}}\right)^{2}(1-\mu^2)\right]}\right\}\left\{1-\exp{\left[-\left(\frac{k}{k_{\parallel,\text{min}}}\right)^{2}\mu^2\right]}\right\}.
\end{equation}

With one final assumption that the dominant mode of redshift-space distortion is linear growth of structure, we obtain
\begin{equation}
    \tilde{P}(k,\mu)=\eta^2\left[\avg{Tb}^2\left(1+\frac{\Omega_m(z)^{0.55}}{b}\mu^2\right)^2P_m(k)+P_s\right]W_r(k,\mu)W_V(k,\mu),\label{eq:Pconv}
\end{equation}

As $\tilde{P}(k,\mu)$ is even in $\mu\in(-1,1)$,
$\tilde{P}(k) = (1/2)\int_{-1}^1d\mu\,\tilde{P}(k,\mu) =  \int_0^1d\mu\,\tilde{P}(k,\mu).$ So for each CO transition, given $\sigma_\perp\propto\theta_B$ and $\sigma_\parallel$ defined as above for all lines,
\begin{equation}
    \tilde{P}_{J\times J}(k)=\eta_J^2\int_0^1d\mu\,\left[\avg{Tb}_J^2\left(1+\frac{\Omega_m(z)^{0.55}}{b_J}\mu^2\right)^2P_m(k)+P_{s,J}\right]W_r(k,\mu)W_V(k,\mu).
\end{equation}

\begin{figure}
    \centering
    \includegraphics[width=0.5\linewidth]{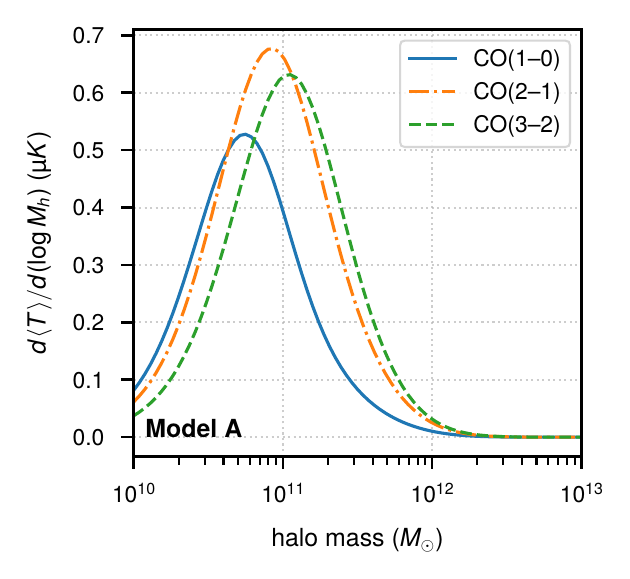}\includegraphics[width=0.5\linewidth]{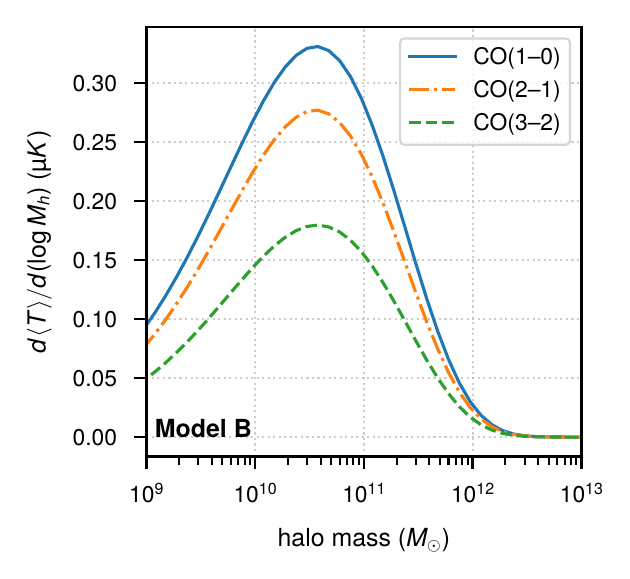}
    \caption{Differential contribution to the average line brightness temperature $\avg{T}$ per logarithmic halo mass bin for each CO transition considered in this work, for Model A (left) and Model B (right).}
    \label{fig:dTdM}
\end{figure}

As an illustration of the halo masses that tend to dominate the overall comoving line luminosity density that determines the signal, \autoref{fig:dTdM} shows the integrand of~\autoref{eq:Tavg} for the three CO transitions considered. In both Model A and Model B, halos of mass $\gtrsim10^{12}\,M_\odot$ are too rare to contribute significantly to the predicted clustering signal, with the differential brightness temperature peaking for halos of mass $\sim10^{10}$--$10^{11}\,M_\odot$.

Cross power spectra calculations are largely similar. The cross shot noise amplitude is calculated as
\begin{equation}
    P_{s,J\times J'} = \frac{(1+z)^4}{H^2(z)}\int dM_h\,\frac{dn}{dM_h}\,L'_J(M_h)L'_{J'}(M_h),\label{eq:Pshot_cross}
\end{equation}
and based on this we may find the total cross-correlation pseudo-power spectrum as
\begin{align}
    \tilde{P}_{J\times J'}(k)&=\eta_J\eta_{J'}\int_0^1d\mu\,\left[\avg{Tb}_J\avg{Tb}_{J'}\left(1+\frac{\Omega_m(z)^{0.55}}{b_J}\mu^2\right)\left(1+\frac{\Omega_m(z)^{0.55}}{b_{J'}}\mu^2\right)P_m(k)\right.\nonumber\\&\hspace{6cm}\left.+P_{s,J\times J'}\right]W_r(k,\mu)W_V(k,\mu).
\end{align}

\subsection{Noise: COMAP survey parameters at reionisation}
\label{sec:noise}
For survey parameters, we borrow the COMAP-EoR and COMAP-ERA concepts proposed in a COMAP Collaboration paper~\citep{Breysse22}. These assume receivers operating at 30 GHz with 19 spectrometers per dish and system temperature $T_\text{sys}=44$ K, and at 15 GHz with 38 spectrometers (19 dual-polarisation feeds) per dish and $T_\text{sys}=22$ K. We follow the original paper in assuming 7000 dish-hours in the Ka band and 29000 dish-hours in the Ku band for the COMAP-EoR survey, and a significantly higher 57000 dish-hours in the Ka band and 110000 dish-hours in the Ku band for the COMAP-ERA survey. In all cases, we make the highly approximate assumption of uniform noise achieved across $N_F=3$ survey fields, each of size $\Omega_F=4$\,deg$^2$.

Additionally, we consider here a COMAP `Triple Deluxe' concept, which adds a Q band component to the COMAP-ERA proposal. We conservatively assume 38 spectrometers (19 dual-polarisation feeds) with $T_\text{sys}=60$\,K for the system temperature\footnote{60\,K is a highly conservative value, principally as a hedge against poor atmospheric conditions. Compare, for example, the achieved system temperature of $\approx38$ K for the Q/U Imaging ExperimenT (QUIET) Q-band coherent amplifier array~\citep{QUIETRX} (albeit from the Atacama desert, a particularly dry site with median precipitable water vapour of 1.2 mm during QUIET's Q-band observations), or the expected system temperature of $\approx45$ K for the Next-generation Very Large Array (ngVLA) 41 GHz receivers~\citep{ngVLA}.} and only 30000 dish-hours achieved on sky in the same fields.

In all cases, the number of dish-hours $t_\text{obs}$ and the number of spectrometers per dish $N_{s/d}$ provides the total integration time per sky pixel of solid angle $\Omega_\text{px}$:
\begin{equation}
    t_\text{px} = \frac{N_{s/d}t_\text{obs}\Omega_\text{px}}{\Omega_F}.
\end{equation}
The noise level per voxel is then given simply by the radiometer equation, with an additional dependence on the bandwidth $\delta\nu$ per frequency channel:
\begin{equation}
    \sigma_N=\frac{T_\text{sys}}{\sqrt{\delta\nu\,t_\text{px}}}.
\end{equation}
Given the comoving volume per voxel $V_\text{vox}$, we then obtain the contribution of noise to the observed power spectrum as
\begin{equation}
    P_N=\sigma_N^2V_\text{vox}.
\end{equation}
Where necessary, we assume $\delta\nu=2$\,MHz and $\Omega_\text{px}=\theta_B/\sqrt{8\ln{2}}$. However, in reality the factors of $\delta\nu^{-1}$ and $\Omega_\text{px}^{-1}$ in $\sigma_N^2$ will cancel the factors of $\delta\nu$ and $\Omega_\text{px}$ that must figure into the calculation of $V_\text{vox}$. Therefore, the calculation of $P_N$ is ultimately agnostic to voxelisation.

Beyond power spectrum uncertainties from sample variance and radiometer noise, we must consider the contribution of interloper line emission to the 30 GHz and 45 GHz observations. Suppose that interloper line emission in CO($J_i\to J_i-1$) from some redshift $z\sim z_{Ji}$ is described by the power spectrum $P_J(k_\parallel,k_\perp)$ in its original comoving frame. In the $z=z_\text{target}\sim7$ observation, this interloper component will present with modes scaled in the transverse dimensions by
\begin{equation}
    \alpha_\parallel=\frac{H(z_\text{target})}{H(z=z_{Ji})}\frac{1+z_{Ji}}{1+z_\text{target}},
\end{equation}
and along the line of sight by
\begin{equation}
    \alpha_\perp=\frac{\chi(z_{Ji})}{\chi(z_\text{target})},
\end{equation}
so that the apparent power spectrum is
\begin{equation}
    P_{Ji,\text{proj}}(k_\parallel,k_\perp) = \alpha_\parallel^{-1}\alpha_\perp^{-2}P_{Ji}(k_\parallel/\alpha_\parallel,k_\perp/\alpha_\perp),
\end{equation}
as a function of the line-of-sight and transverse components $(k_\parallel,k_\perp)$ of the wavevector in the projected $z\sim7$ comoving frame. This projected power spectrum may then be subjected to the same resolution and volume window functions as the primary target power spectra, as well as the appropriate factor of $\eta_J^2$, to yield the projected spherically averaged pseudo-power spectrum $\tilde{P}_{Ji,\text{proj}}(k)$. (Note that the definition of the original $P_{Ji}(k_\parallel,k_\perp)$ would cover redshift-space distortions from linear structure growth.)

We consider all interlopers to follow a luminosity function $\Phi(L')$ parameterised as a Schechter function with a smooth but rapid cutoff at low luminosities, and tuned to mimic the luminosity function of~\cite{Li16}.\footnote{P.~C.~Breysse, priv.~comm.} Given that the interloper emission is not a primary focus of this paper and largely manifests as a source of additional uncertainty in the measurement of reionisation-epoch signals, we consider it appropriate to apply this level of modelling across both the $z\sim3$ CO(1--0) interloper in the 30 GHz observation and the two additional interloper components that would manifest in the 45 GHz observation. \autoref{sec:interloper} provides further details including a favourable comparison against current observational constraints on CO luminosity functions at $z\sim1$--4.

With the power spectrum for thermal noise and any interlopers, the total observed auto pseudo-power spectrum from an observation targeting CO($J\to J-1$) at $z\sim7$ becomes
\begin{equation}
    \tilde{P}_{J\times J,\text{tot}}(k) = P_N + \tilde{P}_{J\times J}(k) + \sum_{\forall J_i:1\leq J_i<J}\tilde{P}_{Ji,\text{proj}}(k).
\end{equation}
The expected values of all cross pseudo-power spectra remain unaffected (i.e., $\tilde{P}_{J\times J',\text{tot}}(k) = \tilde{P}_{J\times J'}(k)$ for $J\neq J'$) as all assumed sources of noise and interloper emission are independent between the different observations being cross-correlated.

At last we are ready to calculate the uncertainties for auto and cross power spectra. We assume Gaussianity, meaning diagonal covariances within each observable so that no correlations exist between different wavenumber bins. Within each wavenumber bin of width $\Delta k$ (not necessarily uniform) we expect to observe a number of independent modes equal to
\begin{equation}
    N_m=N_F\times L_\perp^2L_\parallel\frac{k^2\,\Delta{k}}{4\pi^2}.
\end{equation}
(The total number of modes observed is twice this, but half of the Fourier modes perfectly track the other half for Fourier transforms of real variables.) Then the uncertainty in the auto pseudo-power spectrum at each wavenumber is given simply as
\begin{equation}
    \sigma_{J\times J}(k) = \frac{\tilde{P}_{J\times J,\text{tot}}(k)}{\sqrt{N_m(k)}}.
\end{equation}
Most generally, the covariance between any of the pseudo-power spectra will be given as
\begin{equation}
    \operatorname{cov}\left[\tilde{P}_{J_1\times J_1'}(k),\tilde{P}_{J_2\times J_2'}(k)\right]=\frac{\tilde{P}_{J_1\times J_2,\text{tot}}(k)\tilde{P}_{J_1'\times J_2',\text{tot}}(k)+\tilde{P}_{J_1\times J_2',\text{tot}}(k)\tilde{P}_{J_2\times J_1',\text{tot}}(k)}{2N_m(k)},
\end{equation}
owing to the assumption of Gaussian random fields and the consequent ability to express fourth-order moments purely in terms of second-order moments (cf., e.g., Appendix A of~\cite{White09} for more explicit consideration).
This recovers the expression for the variance of the cross pseudo-power spectrum, given by~\cite{Breysse22} as\begin{equation}
    \sigma^2_{J\times J'}(k) = \frac{1}{2}\left(\frac{\tilde{P}_{J\times J'}(k)}{N_m(k)}+\sigma_{J\times J}(k)\sigma_{J'\times J'}(k)\right),
\end{equation}
and the covariance between auto and cross pseudo-power spectra, given by~\cite{Breysse22} as
\begin{equation}
    \operatorname{cov}\left[\tilde{P}_{J\times J}(k),\tilde{P}_{J\times J'}(k)\right] = \frac{\tilde{P}_{J\times J,\text{tot}}(k)\tilde{P}_{J\times J'}(k)}{N_m(k)},
\end{equation}
given $J\neq J'$. However, the most general expression also allows us to consider covariances between different cross pseudo-power spectra. For example, the covariance between the 15 GHz--30 GHz auto pseudo-power spectrum ($J_1=1$, $J_1'=2$) and the 15 GHz--45 GHz cross pseudo-power spectrum ($J_2=1$, $J_2'=3$) is given as
\begin{equation}
    \operatorname{cov}\left[\tilde{P}_{1\times2}(k),\tilde{P}_{1\times3}(k)\right] = \frac{\tilde{P}_{1\times1,\text{tot}}(k)\tilde{P}_{2\times3}(k)+\tilde{P}_{1\times2}(k)\tilde{P}_{1\times3}(k)}{2N_m(k)}.
\end{equation}
This does imply, incidentally, that even when simulating an experimental scenario where only the 15 GHz--30 GHz and 15 GHz--45 GHz cross spectra inform parameter recovery, calculation of the 30 GHz--45 GHz cross spectrum is necessary to inform covariances.

For most of this work, we only use wavenumbers in the range of $k\in(0.03,0.5)$\,Mpc$^{-1}$, including when calculating the total signal-to-noise ratio for each observable:
\begin{equation}
    \mathrm{S/N}_{J\times J',\text{tot}}\equiv\sqrt{\sum_{k}\left(\frac{\tilde{P}_{J\times J'}(k)}{\sigma_{J\times J'}(k)}\right)^2}.
\end{equation}
(For a combination of multiple observables, this generalises to the square root of the vector-matrix-vector product between the transposed observable vector, the full covariance matrix, and the observable vector.) We take this step to avoid unduly over-emphasising wavenumbers where in reality miscalibrated beams or other transfer functions may strongly bias measurements. We show observables in some contexts across a wider range of $k\in(10^{-2},10^1)\,$Mpc$^{-1}$, but purely for illustrative purposes.

\subsection{Recovery capability}
To test (albeit in a highly idealised simulation) whether any of the COMAP scenarios described above are capable of recovering the original model parameters, we run a series of Markov chain Monte Carlo (MCMC) exercises using \texttt{emcee}~\citep{emcee}. In addition to model parameters for the $z\sim7$ CO flux calculation, we also explore the posterior distribution for three nuisance parameters describing the 30 GHz interloper emission in CO(1--0) from $z\sim3$. The CO(3--2) auto power spectrum is never used, so we forgo the need to marginalise over six more nuisance parameters for the two interloper line components in the Q band.

We provide fairly constrained Gaussian priors on $\avg{Tb}$, $b$, and $P_\text{shot}$ for the $z\sim3$ CO(1--0) interloper (20\%, 33\%, and 10\% relative uncertainties in each case). This reflects an expectation that external data (from ngVLA surveys, for instance, or even from the COMAP Pathfinder) may constrain parameters for this CO component by the time of COMAP-EoR.

Otherwise, we use broad tophat priors for all model parameters.
\begin{itemize}
    \item For Model A, we allow $\alpha\in(0,2)$, $\log{(M_\text{break}/M_\odot)}\in(10,13)$, $\log{(N_\text{break}/\text{cm}^{-2})}\in(10,22)$, $T_k\in(20,1000)$, and $\log{(n_\text{H2,10}/\text{cm}^{-2})}\in(0,6)$.
    \item For Model B, we allow $\alpha\in(0,2)$, $\log{(M_\text{break}/M_\odot)}\in(9,13)$, $\log{(N_\text{break}/\text{cm}^{-2})}\in(10,22)$, $\log{(T_k/\text{K})}\in(1.4,2.5)$ (corresponding to $T_k/\text{K}\in(25,320)$ approximately), $\log{n_\text{H2}/\text{cm}^{-2}}\in(0,6)$, $\beta\in(1,3)$, $\log{N_\text{MC,brk}}\in(2,10)$, and $\log{(M_\text{min}/M_\odot)}\in(7.5,13)$.
\end{itemize}
Most of these priors are meant to be broad and uninformative. The range of allowed $T_k$ values in Model B reflects an expectation that the dust temperature $T_\text{dust}$ is in the range of 50 to 100 K, and that the kinetic temperature is comparable to this but is driven by slightly different physics to the point where $T_k/T_\text{dust}\in(0.5,3)$. We allow a slightly wider range for Model A to reflect the more phenomenological nature of the model.

\autoref{tab:scenarios} describes the COMAP scenarios considered and the observables used in simulating each scenario. Note in particular that we consider an alternate COMAP-ERA scenario where the Ku band instrumentation is replaced rather than supplemented (as in `Triple Deluxe') with Q band instrumentation, requiring us to fall back on the CO(2--1)--CO(3--2) cross and CO(2--1) auto power spectra. In fact, neither of these observables are used in the ideal `Triple Deluxe' simulation, which leverages the Ku band data to detect the CO(1--0) auto, CO(1--0)--CO(2--1) cross, and CO(1--0)--CO(3--2) cross power spectra, probing the CO SLED with the ground transition as a firm reference point.

Note that we also add a number of idealised scenarios. In two of these, only the 15 GHz or 30 GHz survey occurs but with infinite survey time such that the radiometer noise goes to zero. This results in a noise-free 15 GHz standalone survey in one case, and a noise-free, interloper-limited 30 GHz standalone survey in the other. We also consider a scenario where COMAP-ERA only operates in the Ka band with 110000 dish-hours, but the interloper has somehow been subtracted with perfect knowledge so that its contribution to the uncertainty in the reionisation-epoch CO(2--1) signal may be excluded from the noise budget. In this scenario (and the full `Triple Deluxe' scenario), we do not marginalise over the interloper nuisance parameters as there is no need to do so.

\begin{table}
    \centering
    \begin{tabular}{cccccccp{3.36cm}}
        & & \multicolumn{3}{c}{Ku $\times$ \dots} & \multicolumn{2}{c}{Ka $\times$ \dots} & \\ \multicolumn{2}{c}{Experimental scenario} & Ku & Ka & Q & Ka & Q & Other notes
        \\\hline
        Noise-free & 15 GHz only & \textbf{Y}&---&---&---&---& Infinite survey time \\ 
        Interloper-limited & 30 GHz only & ---&---&---&\textbf{Y}&---& Infinite survey time\\
        COMAP-EoR & 15/30 GHz & \textbf{Y}&\textbf{Y}&---&\textbf{Y}&---& ---\\
        COMAP-ERA & 15 GHz only & \textbf{Y}&---&---&---&---& --- \\ & 30 GHz only & ---&---&---&\textbf{Y}&---& --- \\ & (interloper-free) & ---&---&---&\textbf{Y}&---& $z\sim3$ CO(1--0) line foreground removed from noise budget \\
         & 15/30 GHz & \textbf{Y}&\textbf{Y}&---&\textbf{Y}&---& ---\\
        COMAP-ERA (alt.)
        & 30/45 GHz & ---&---&---&\textbf{Y}&\textbf{Y}& ---\\
        `Triple Deluxe' &15/30/45 GHz & \textbf{Y}&\textbf{Y}&\textbf{Y}&---&---& ---
    \end{tabular}
    \caption{Experimental scenarios considered in this work, and the specific auto and cross spectra included in mock parameter recovery for each scenario. COMAP-ERA (COMAP-EoR) scenarios assume 57000 (7000) dish-hours at 15 GHz and 110000 (29000) dish-hours at 30 GHz. Both the alternate COMAP-ERA and `Triple Deluxe' scenarios assume accrual of 30000 dish-hours at 45 GHz. The noise budget for the 30 GHz observation includes both contributions from the radiometer noise and the CO(1--0) interloper emission from $z\sim3$, unless noted otherwise in the last column.}
    \label{tab:scenarios}
\end{table}

\subsubsection{Streamlining evaluation: a Gaussian process emulator for Model B}
\label{sec:emulator}
Running MCMC exercises requires as many as millions of evaluations of the outputs of our CO model. For instance, for the Model A MCMC exercises, we iterate 48 walkers over 55865 steps, discarding the 16000 steps as burn-in. This means we calculate the CO observables a total of almost 2.7 million times. While the Radex-based halo model calculation is relatively fast compared to a full hydrodynamical simulation, it still demands enough computational resources for each MCMC exercise to take multiple days of continuous wall time.

There is little leeway around this for Model A, due to the relatively wide range of column densities and number densities that may exist across the halo population. However, for Model B, the Radex calculation is done not per halo but per cloud, so that we have a narrower expectation for the range of plausible environmental conditions. Motivated by this, we design an `emulator' to interpolate the Radex calculation of CO fluxes per cloud for a given set of parameters, in lieu of a simple lookup table.

We use \texttt{scikit-learn}~\citep{scikit-learn} to generate a Gaussian process regression against Radex-generated fluxes in CO(1--0), CO(2--1), and CO(3--2), given inputs spanning $T_k\in(25,320)$ K, $n_\text{H2}\in(10^0,10^6)$ cm$^{-3}$, and $N_\text{CO}\in(10^6,10^{20})$ cm$^{-2}$ in a grid of $25\times25\times29$ points. The model uses the linear combination of a constant kernel, a radial basis function kernel, and a rational quadratic kernel. We allow \texttt{scikit-learn} to optimise the kernel hyperparameters given the provided inputs and outputs.

After fully optimising the regression, we verify the performance of our Gaussian process emulator with 27000 randomly drawn values of $\{T_k,n_\text{H2},N_\text{CO}\}$ in the range. Calculation of these outputs took roughly 12 seconds with the Gaussian process emulator, versus 220 seconds with Radex. Relative errors for fluxes were contained within 5\% in the overwhelming majority of cases, with the 99.9\% CI for GP prediction being $1.00_{-0.03}^{+0.04}$ times the actual flux value for each line. The accuracy is best towards the middle of this column density range, but even a run of 8000 randomly drawn values restricted to $N_\text{CO}\in(10^{15},10^{20})$ cm$^{-2}$ found $1.00_{-0.03}^{+0.05}$ for $J=1$ and $J=3$, $1.00\pm0.03$ for $J=2$. Even extreme outliers in this run fell within 20\% of truth. The acceptable accuracy of the emulator as a surrogate for Radex and its order-of-magnitude improvement in calculation speed allows use of more and larger MCMCs for Model B, using 64 walkers over 86555 steps, discarding 30969 steps as burn-in.

Note that we have only trained our emulator up to column density values of $N_\text{CO}=10^{20}$ cm$^{-2}$. The reason lies in the typical column density values for our fiducial incarnation of Model B, where halos of $M_h\lesssim10^{11}\,M_\odot$ hosting clouds with $N_\text{CO}\sim10^{18}$ cm$^{-2}$ dominate the signal. Values much higher may be unphysical or violate assumptions surrounding our model. For one, Radex itself limits the user to evaluating CO fluxes for $N_\text{CO}\in(10^5,10^{25})$ cm$^{-2}$. But furthermore, if we believe~\autoref{eq:NCO_modelB}, then a value of $N_\text{CO}=10^{20}$ cm$^{-2}$ in the average molecular cloud requires 100\% of carbon to be locked up in CO with, e.g., $n_\text{H2}=10^5\,$cm$^{-3}$ and $Z'=Z_\odot$, or $n_\text{H2}=10^6$\,cm$^{-3}$ and $Z'=0.2Z_\odot$---fairly extreme expectations for the nascent molecular clouds of $z\sim7$. Given all of this, during the MCMC we clip column densities to the range covered by the emulator before evaluating line fluxes.

By reverting to a prediction for lower column densities than the model would otherwise predict, this truncation could bias us subtly towards models that favour higher CO column densities than `truth', i.e., higher values of $N_\text{CO,brk}$ or $n_\text{H2}$. In practice, we find that the posteriors for the multi-band COMAP scenarios tend not to clip into the the upper limit, and certainly not into the lower limit. In fact, the recovered distribution of both $N_\text{CO,brk}$ and $n_\text{H2}$ tend to have heavy left tails, not the heavy right tails that one might expect based on the emulator input truncation. We show this and the full posterior distribution for all MCMC exercises in~\autoref{sec:mcmc_full}.

We emphasise that Gaussian process emulators allow highly flexible interpolation of smooth multivariate output from multivariate input, but in some cases require significant tuning of kernel forms and/or have difficulty with highly unsmooth outputs. We also note that the emulator used in this work does not fully explore the possibilities of Gaussian processes. For instance, the emulator here is agnostic to correlations between CO lines, which could be used to the advantage of an emulator with an appropriate cross-covariance model (see, e.g., the review by~\cite{GentonKleiber15}---although see also~\cite{ZhangCai15}).

\section{Results}
\label{sec:results}
We first discuss in~\autoref{sec:snr} the raw detectability of the observables given our CO emission models. Following this, we consider in~\autoref{sec:mcmc_results} whether the various single- and multi-band surveys simulated are fundamentally capable of recovering ISM properties.
\subsection{Observables and signal-to-noise}
\label{sec:snr}
\begin{figure}
    \centering
    \includegraphics[width=0.86\linewidth]{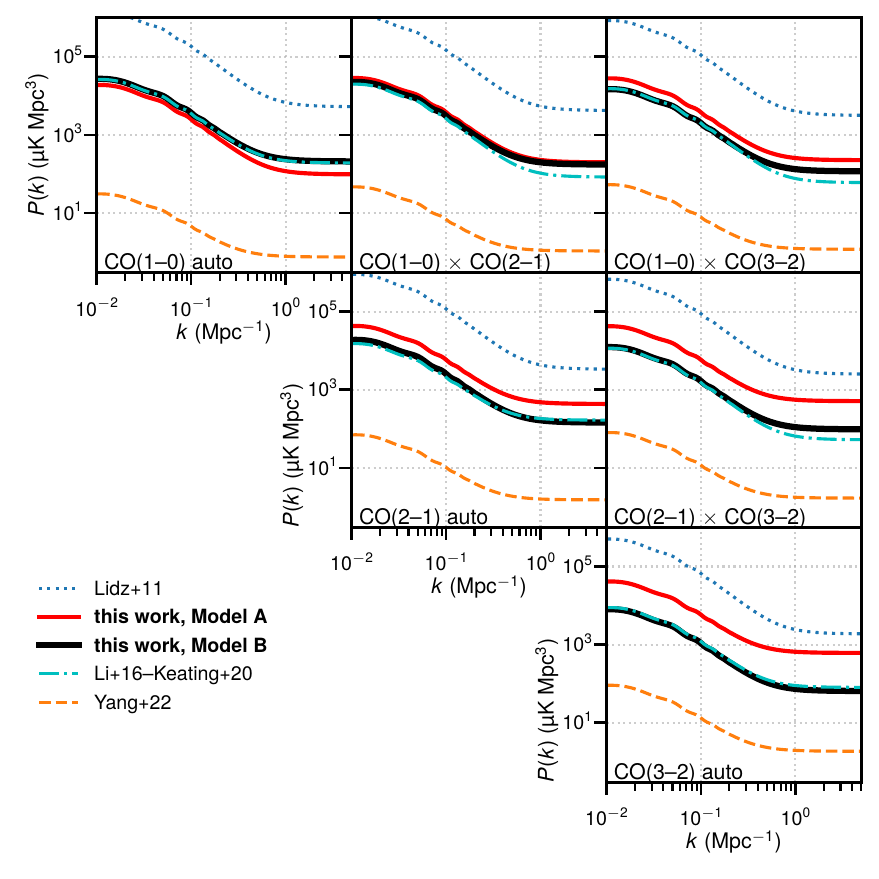}
    \caption{Predictions for CO auto and cross power spectra based on Model A (red solid curves) and Model B (black solid curves) from this work. We also show a number of models from existing literature as considered in~\cite{Breysse22}, representing the range of predictions; see the main text for details.}
    \label{fig:models}
\end{figure}
\autoref{fig:models} shows all of our predicted CO auto and cross power spectra based on both Model A and Model B. While some of our assumptions around both models are somewhat optimistic, we note that our predictions fall within an order of magnitude of the so-called `Li+16--Keating+20' model~\citep{Li16,mmIME-ACA}, which falls in the middle of the pack of models considered by~\cite{Breysse22}. Given that this model combines a number of fairly straightforward scaling relations between halo mass, star-formation rate, infrared luminosity, and CO line luminosities drawn from a number of numerical and observational works~\citep{Behroozi13a,Behroozi13b,Kamenetzky16}, it is not an unreasonable model for us to track. For additional reference, we also reproduce the two models at the extremes of the pack considered by~\cite{Breysse22}: the `Lidz+11' model from~\cite{Lidz11} featuring a simple linear scaling relation between halo mass and CO luminosity, and the `Yang+22' model from~\cite{Yang22} featuring scaling relations calibrated based on semi-empirical models applied to cosmological simulations. We refer the reader to the individual papers, including~\cite{Breysse22}, for further details on the models not introduced in this work, which we have recalculated at $z=6.7$ for~\autoref{fig:models}.

\begin{figure}
    \includegraphics[width=0.5\linewidth]{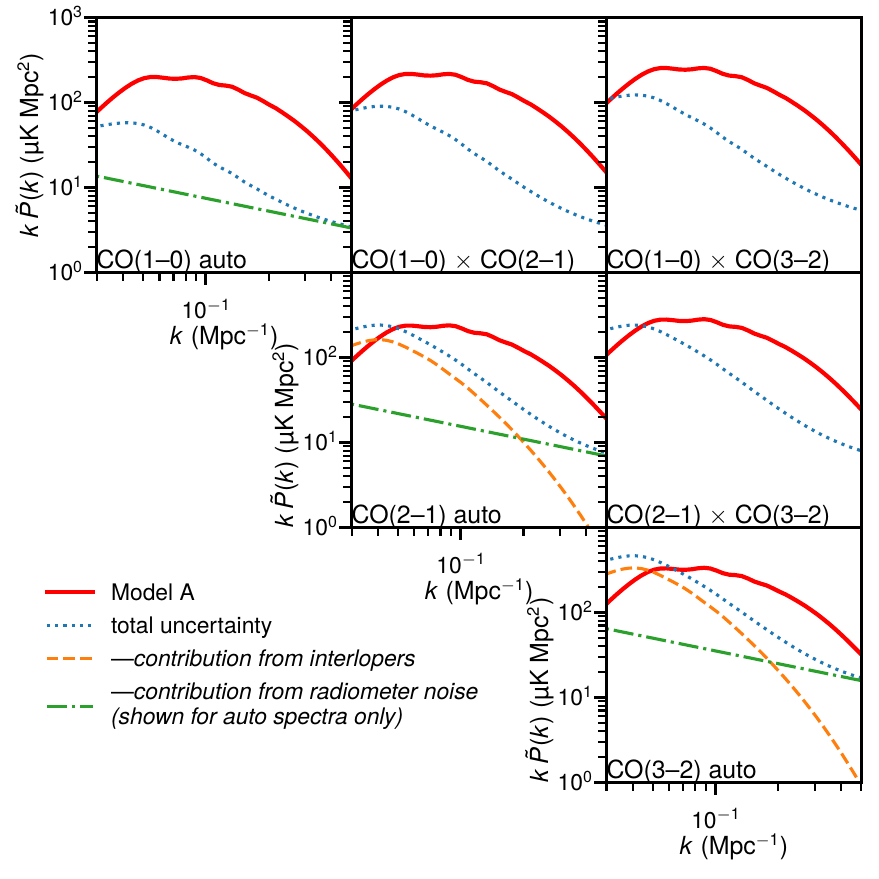}\includegraphics[width=0.5\linewidth]{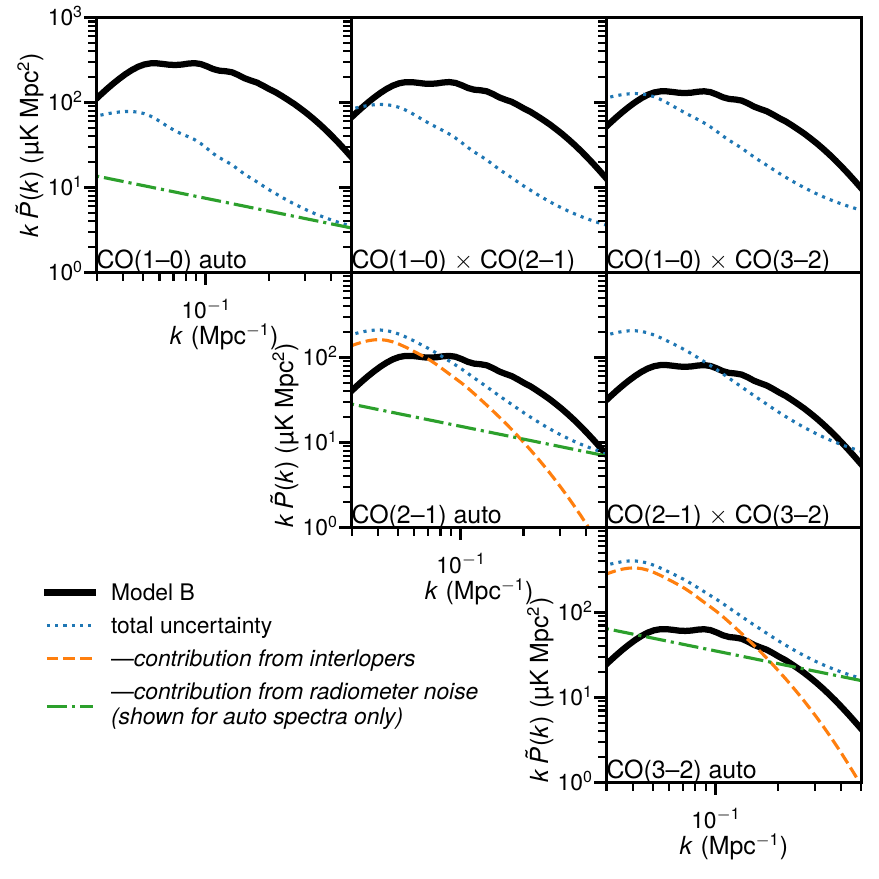}
    \caption{Predictions for CO auto and cross pseudo power spectra given Model A (left panels) and Model B (right panels), alongside the expected uncertainty on the power spectra evaluated for $k$-bins of size $\Delta[\log{(k\,\text{Mpc})}]=0.03$. We show the sensitivities expected for the COMAP-ERA or COMAP Triple Deluxe scenarios; for COMAP-EoR, the radiometer noise will be worse for the CO(1--0) auto power measurement by a factor of 8, and for CO(2--1) by a factor of 4.}
    \label{fig:models_noise}
\end{figure}

Based on the parameters and equations laid out in~\autoref{sec:noise}, we calculate the uncertainties for each observable and the signal-to-noise ratio obtained in each survey scenario. \autoref{fig:models_noise} shows all pseudo-power spectra and sensitivities for the COMAP `Triple Deluxe' scenario, including for the unused CO(3--2) auto power spectrum. Under Model B, the entirely sub-thermal CO excitation predicted renders the two higher transitions dimmer but not out of reach of a detection, especially in cross-correlation against CO(1--0). However, CO(2--1) and CO(3--2) emission are certainly easier to detect under the more optimistic Model A.

\begin{table}
    \centering
    \begin{tabular}{cccccccc}
        & & \multicolumn{3}{c}{Ku $\times$ \dots} & \multicolumn{2}{c}{Ka $\times$ \dots} & \\ \multicolumn{2}{c}{Experimental scenario} & Ku & Ka & Q & Ka & Q & Total signal-to-noise
        \\\hline
        Interloper-limited & 30 GHz only & ---&---&---&76&---& 76\\
        COMAP-EoR & 15/30 GHz & 11&14&---&10&---& 19\\
        COMAP-ERA & 15 GHz only & 51&---&---&---&---& 51 \\
         & 15/30 GHz & 51&44&---&21&---& 60\\
        COMAP-ERA (alt.)
        & 30/45 GHz & ---&---&---&21&25& 31\\
        `Triple Deluxe' &15/30/45 GHz & 51&44&36&---&---& 63
    \end{tabular}
    \caption{Predictions for signal-to-noise ratios both for individual observables and in total for selected scenarios described in~\autoref{tab:scenarios}, assuming Model A. COMAP-ERA (COMAP-EoR) scenarios assume 57000 (7000) dish-hours at 15 GHz and 110000 (29000) dish-hours at 30 GHz. `Triple Deluxe' assumes 30000 dish-hours at 45 GHz, on top of COMAP-ERA sensitivities at 15 GHz and 30 GHz. The noise budget for the 30 GHz observation includes both contributions from the radiometer noise and the CO(1--0) interloper emission from $z\sim3$, unless noted otherwise in the last column.}
    \label{tab:SNR_modelA}
\end{table}

\begin{table}
    \centering
    \begin{tabular}{cccccccc}
        & & \multicolumn{3}{c}{Ku $\times$ \dots} & \multicolumn{2}{c}{Ka $\times$ \dots} & \\ \multicolumn{2}{c}{Experimental scenario} & Ku & Ka & Q & Ka & Q & Total signal-to-noise
        \\\hline
        Noise-free & 15 GHz only & 310&---&---&---&---& 310 \\ 
        Interloper-limited & 30 GHz only & ---&---&---&35&---& 35\\
        COMAP-EoR & 15/30 GHz & 16&12&---&4&---& 19\\
        COMAP-ERA & 15 GHz only & 65&---&---&---&---& 65 \\ & 30 GHz only & ---&---&---&10&---& 10\\ & (interloper-free) & ---&---&---&20&---& 20 \\
         & 15/30 GHz & 65&35&---&10&---& 67\\
        COMAP-ERA (alt.)
        & 30/45 GHz & ---&---&---&10&7& 12\\
        `Triple Deluxe' &15/30/45 GHz & 65&35&19&---&---& 70
    \end{tabular}
    \caption{Predictions for signal-to-noise ratios both for individual observables and in total for the experimental scenarios described in~\autoref{tab:scenarios}, assuming Model B. COMAP-ERA (COMAP-EoR) scenarios assume 57000 (7000) dish-hours at 15 GHz and 110000 (29000) dish-hours at 30 GHz. `Triple Deluxe' assumes 30000 dish-hours at 45 GHz, on top of COMAP-ERA sensitivities at 15 GHz and 30 GHz. The noise budget for the 30 GHz observation includes both contributions from the radiometer noise and the CO(1--0) interloper emission from $z\sim3$, unless noted otherwise in the last column.}
    \label{tab:SNR_modelB}
\end{table}

We confirm this quantitatively with the calculated signal-to-noise values shown in~\autoref{tab:SNR_modelA} and~\autoref{tab:SNR_modelB}. Given either model, however, the two-band COMAP-EoR scenario achieves a significant detection of CO emission with total signal-to-noise of 17--18 across all observables. COMAP-ERA and `Triple Deluxe' scenarios also achieve total signal-to-noise near or above 60.
\subsection{Recovery capability}
\label{sec:mcmc_results}
\begin{figure}
    \centering
    \includegraphics[width=0.5\linewidth]{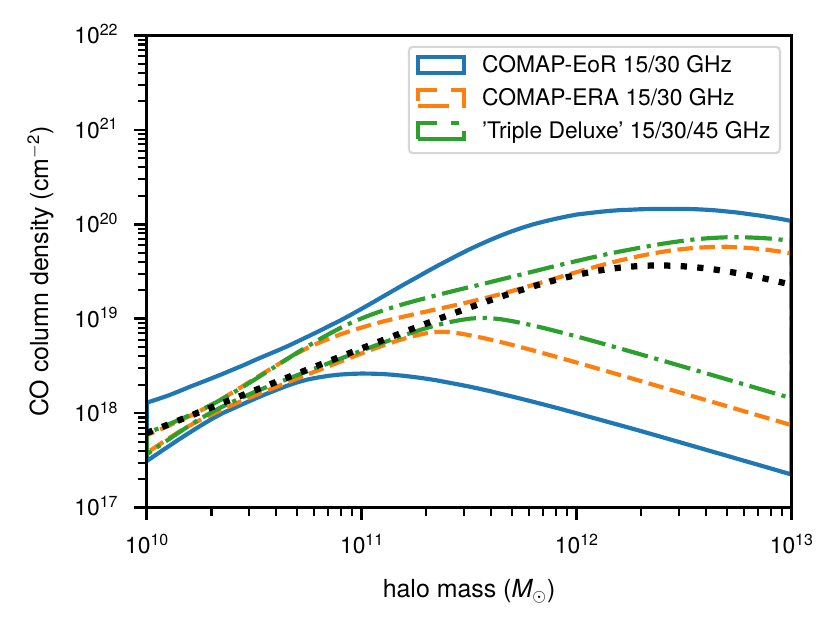}\includegraphics[width=0.5\linewidth]{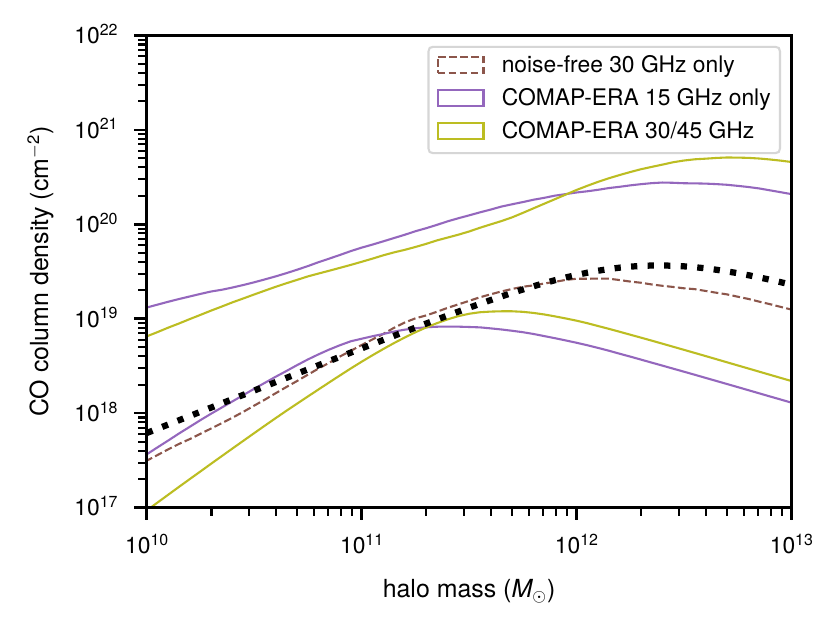}
    
    \includegraphics[width=0.5\linewidth]{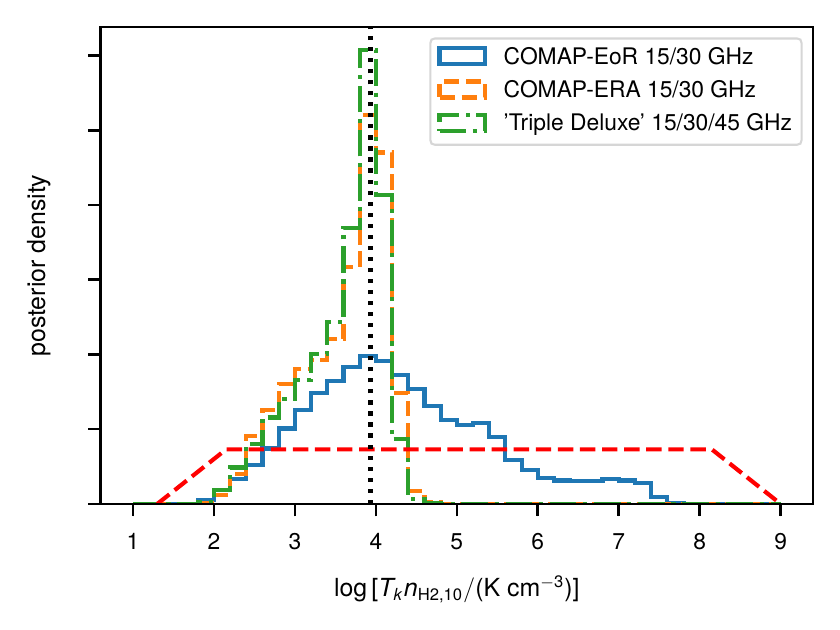}\includegraphics[width=0.5\linewidth]{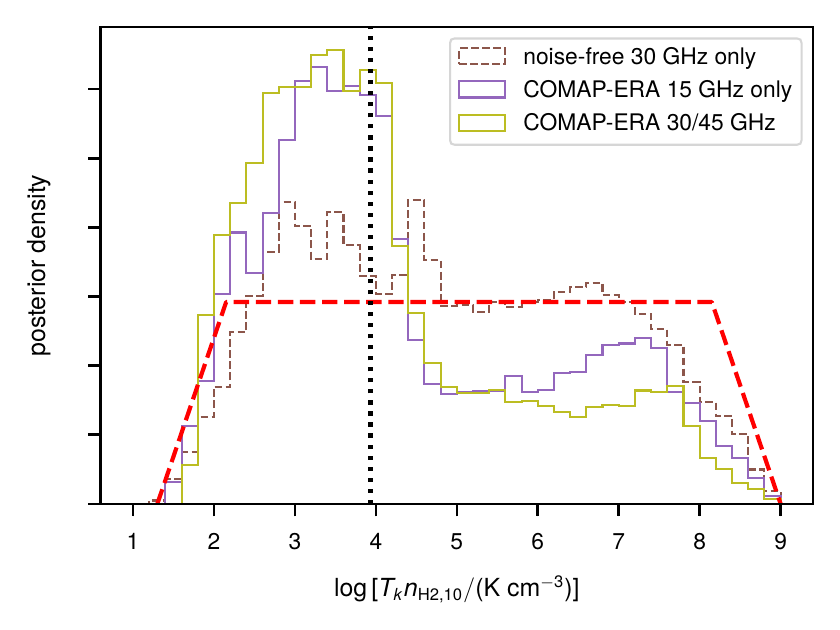}
    \caption{Predicted constraints, given Model A, on the CO column density per halo (upper panels) and the H$_2$ thermal pressure derived from the product of kinetic temperature and number density for a dark matter halo of mass $10^{10}\,M_\odot$ (lower panels). The left and right panels show different experimental scenarios as indicated in the legends. Also shown for reference are the true parameter values for Model A (black dotted lines), as well as the distribution of H$_2$ thermal pressure values implied by the MCMC priors alone (red dashed lines).}
    \label{fig:modelA}
\end{figure}

\begin{table}
    \centering
    \begin{tabular}{cccccc}
        & &$T_k$ (K)&$\log{\frac{n_\text{H2,10}}{\text{cm}^{-3}}}$ &   \multicolumn{2}{c}{$\log{\frac{T_kn_\text{H2,10}}{\text{K\,cm}^{-3}}}$}\\[2pt] \multicolumn{2}{c}{Experimental scenario}& 95\% LL & 95\% UL & 68\% CI & 90\% CI
        \\\hline
        \multicolumn{2}{c}{\emph{True values}} & \emph{86} & \emph{2.00} & \multicolumn{2}{c}{\emph{3.93}} \\
        \multicolumn{2}{c}{\emph{Prior percentiles}} & $\it > 24$ & $\it < 5.70$ & \emph{3.10--7.14} & \emph{(2.29--7.95)}\\
        Interloper-limited & 30 GHz only & ${>24}$ & $<5.65$ & 2.92--7.05 &  (2.28--7.93)\\
        COMAP-EoR & 15/30 GHz & $>27$ & ${<5.18}$ & {\bf 3.18--5.44} & {\bf (2.66--6.65)} \\
        COMAP-ERA & 15 GHz only & $>31$ & $<5.63$ & 2.72--6.76 & (2.14--7.74) \\
         & 15/30 GHz & ${\bf >44}$ & ${\bf <2.31}$ & {2.96--4.09} & {\bf (2.56--4.23)}\\
        COMAP-ERA (alt.)
        & 30/45 GHz & ${\bf >50}$ & ${<5.46}$ & {2.58--6.01} &  (2.08--7.55)\\
        `Triple Deluxe' &15/30/45 GHz & ${\bf >42}$ & ${\bf <2.37}$ & {2.99--4.04} &  {\bf (2.55--4.17)}
    \end{tabular}
    \caption{Predicted constraints, given Model A, on kinetic temperature $T_k$, H$_2$ number density $n_\text{H2,10}$ in dark matter halos of mass $10^{10}\,M_\odot$, and the H$_2$ thermal pressure derived from the product of the two. Also shown for reference are the true parameter values for Model A and the percentiles implied solely by tophat priors bounding the MCMC posterior. Values in bold type indicate shifts away from the prior percentiles by more than 10\% of the 90\% width of the prior distribution (i.e., by more than 0.15 dex for $T_k$, 0.54 dex for $n_\text{H2,10}$, and 0.57 dex for $T_kn_\text{H2,10}$), unless either bound of the posterior interval falls beyond the corresponding bound of the prior.}
    \label{tab:modelA_ci}
\end{table}

\autoref{fig:modelA} and~\autoref{tab:modelA_ci} show the fundamental recovery capabilities under Model A of the experimental scenarios considered in~\autoref{tab:SNR_modelA}. The multi-transition survey simulations that include the Ka band all appear to significantly outperform the single-transition survey simulations in recovering the relation between halo mass and CO column density, particularly at the halo masses ($\lesssim10^{12}\,M_\odot$) that actually dominate the signal. The multi-band COMAP-EoR scenario, despite achieving around roughly one-third of the total signal-to-noise of either single-band survey scenario, recovers the CO column density at low halo masses comparably if not better. COMAP-ERA and `Triple Deluxe' simulations improve recovery of the relation at these masses and extends meaningful constraints to halos of somewhat higher masses, which COMAP-EoR is able to constrain less owing to their expected relative scarcity. Interestingly, the alternate COMAP-ERA scenario forgoing 15 GHz in favour of 30/45 GHz does not improve on the single-band scenarios.

The story is largely similar for the ISM parameters of kinetic temperature, H$_2$ number density, and H$_2$ thermal pressure (the product of the first two parameters). In particular, the `Triple Deluxe' scenario is capable of noticeably constraining all three parameters beyond the limits of the prior distribution. The CO(1--0) emission alone is incapable of constraining any of these variables too much beyond the priors, while the higher transitions by themselves can bound the kinetic temperature but have trouble placing limits on the number density and pressure without the ground transition. The `Triple Deluxe' constraint on the thermal pressure of  $\log{[T_kn_\text{H2}/(\text{K\,cm}^{-3})]}=2.99$--4.04 (68\% CI, for a half-width of 0.53 dex) is somewhat skewed towards lower values as is apparent from the heavy left tail of the posterior distribution shown in~\autoref{fig:modelA}, but the constraint is fully consistent with the ground truth of $\log{[T_kn_\text{H2}/(\text{K\,cm}^{-3})]}=3.93$, and we also see in~\autoref{fig:modelA} that the maximum \emph{a posteriori} value is close to this `true' value.

\begin{figure}
    \centering
    \includegraphics[width=0.5\linewidth]{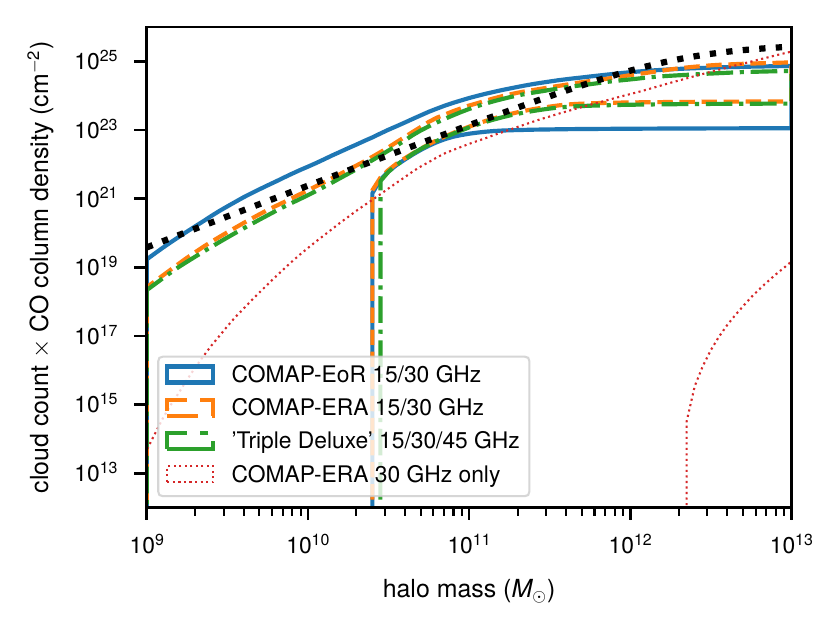}\includegraphics[width=0.5\linewidth]{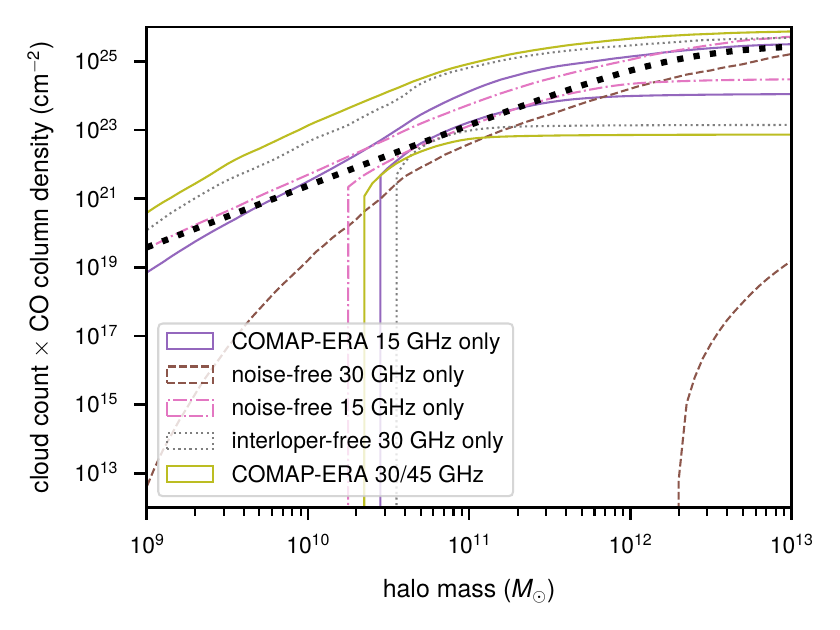}
    
    \includegraphics[width=0.5\linewidth]{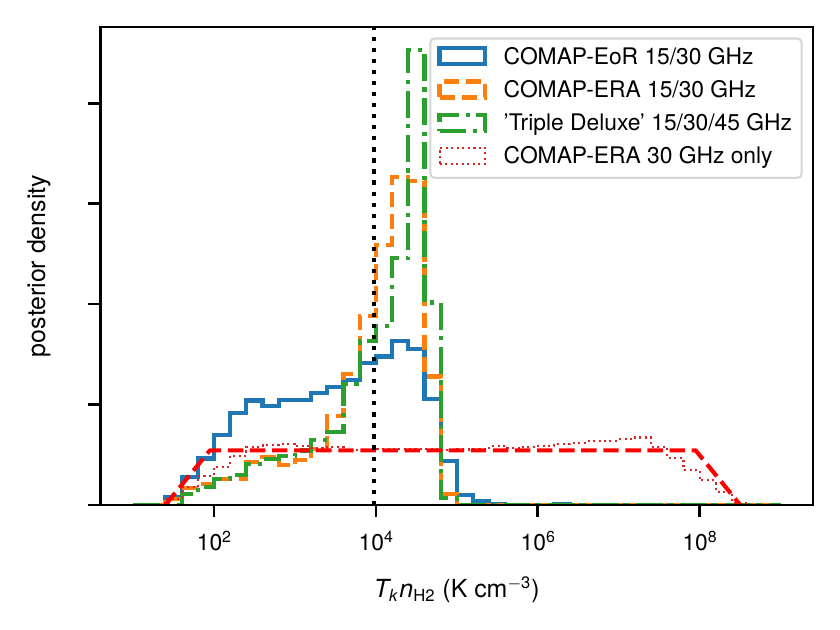}\includegraphics[width=0.5\linewidth]{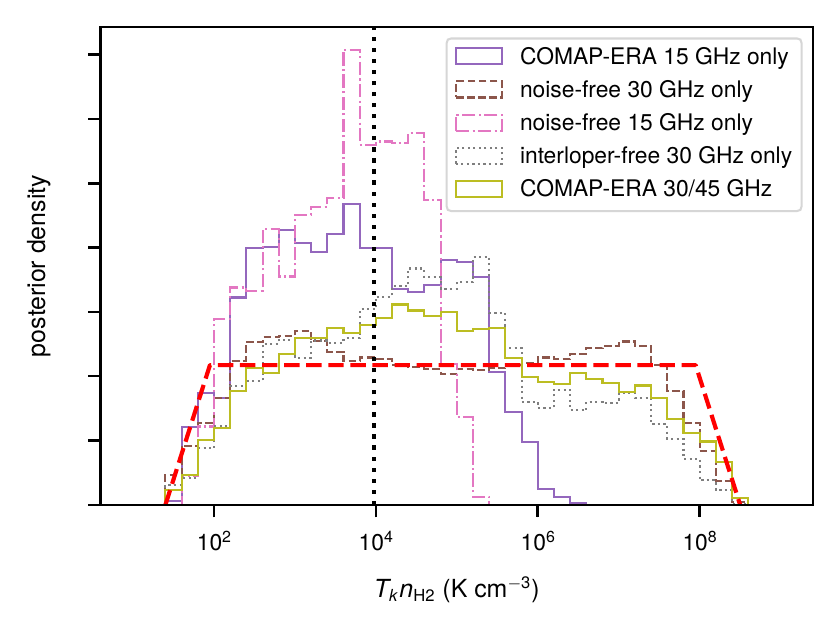}
    \caption{Predicted constraints, given Model B, on the product of the CO column density and cloud count per halo (upper panels) and the H$_2$ thermal pressure derived from the product of kinetic temperature and number density (lower panels). The left and right panels show different experimental scenarios as indicated in the legends. Also shown for reference are the true parameter values for Model B (black dotted lines), as well as the distribution of H$_2$ thermal pressure values implied by the MCMC priors alone (red dashed lines).}
    \label{fig:modelB}
\end{figure}

\begin{table}
    \centering
    \begin{tabular}{cccccc}
        & &$T_k$ (K)&$\log{\frac{n_\text{H2}}{\text{cm}^{-3}}}$ &   \multicolumn{2}{c}{$\log{\frac{T_kn_\text{H2}}{\text{K\,cm}^{-3}}}$}\\[2pt] \multicolumn{2}{c}{Experimental scenario}& 95\% LL & 95\% UL & 68\% CI & 90\% CI
        \\\hline
        \multicolumn{2}{c}{\emph{True values}} & \emph{86} & \emph{2.042} & \multicolumn{2}{c}{\emph{3.98}} \\
        \multicolumn{2}{c}{\emph{Prior percentiles}} & $\it > 29$ & $\it < 5.70$ & \emph{2.94--6.98} & \emph{(2.25--7.67)}\\
        Noise-free & 15 GHz only & ${\bf >39}$ & ${\bf <2.80}$ & {2.60--4.49} & {(2.14--4.81)}  \\ 
        Interloper-limited & 30 GHz only & ${>28}$ & $<5.72$ & 2.78--7.01 &  (2.12--7.68)\\
        COMAP-EoR & 15/30 GHz & $>29$ & ${\bf <2.88}$ & {2.48--4.44} & {(2.03--4.73)} \\
        COMAP-ERA & 15 GHz only & $>33$ & ${\bf <3.47}$ & {2.57--5.04} & {(2.08--5.50)} \\ & 30 GHz only & $>28$ & $<5.75$ & 2.89--7.08 & (2.19--7.73) \\ & (interloper-free) & $>30$ & $<5.54$ & {\bf 2.98--6.44} & (2.29--7.41) \\
         & 15/30 GHz & ${\bf >36}$ & ${\bf <2.74}$ & {\bf 3.27--4.50} & {\bf (2.36--4.65)}\\
        COMAP-ERA (alt.)
        & 30/45 GHz & $>29$ & ${<5.65}$ & 3.02--6.80 &  (2.28--7.69)\\
        `Triple Deluxe'&15/30/45 GHz & ${\bf >42}$ & ${\bf <2.70}$ & {\bf 3.24--4.56} &  {\bf (2.43--4.67)}
    \end{tabular}
    \caption{Predicted constraints, given Model B, on kinetic temperature $T_k$, H$_2$ number density $n_\text{H2}$, and the H$_2$ thermal pressure derived from the product of the two. Also shown for reference are the true parameter values for Model B and the percentiles implied solely by tophat priors bounding the MCMC posterior. Values in bold type indicate shifts away from the prior percentiles by more than 10\% of the 90\% width of the prior distribution (i.e., by more than 0.10 dex for $T_k$ and 0.54 dex for both $n_\text{H2}$ and $T_kn_\text{H2}$), unless either posterior bound falls beyond the corresponding bound of the prior.}
    \label{tab:modelB_ci}
\end{table}

Compared to Model A, we have allowed more parameters to vary for Model B in the mock recovery exercise, and Model B predicts sub-thermal excitation and thus dimmer signals from higher transitions relative to Model A. Nonetheless, we find that many of the findings under Model A apply for Model B, for which we show results across the full range of experimental scenarios in~\autoref{fig:modelB} and~\autoref{tab:modelB_ci}. In particular, in both cases, the typical lower bound of $T_k>42$\,K is within striking distance of the 40--60\,K dust temperature values ascribed to observations of individual objects at $z>6$~\citep{Hashimoto19,Sommovigo20,Ferrara22}, constraining interactions between molecular gas and dust.

One notable difference is that under Model B, tight constraints on the product of cloud count and CO column density per cloud (representing the bulk abundance of CO per halo) do not appear to require multi-band CO LIM. In principle, even surveying CO(1--0) only is capable of achieving constraints equivalent to those achieved by the multi-band versions of COMAP-EoR or COMAP-ERA. However, such results certainly cannot be obtained \emph{without} CO(1--0), barring perfect removal of interloper emission at 30 GHz and 45 GHz.

When it comes to the ISM parameters, we once again find that multi-transition LIM yields the best results, particularly with the inclusion of the CO(1--0) transition. Even with slightly dimmed detection significance for the higher CO transitions, the `Triple Deluxe' simulation recovers the strongest bounds on the kinetic temperature and H$_2$ number density. The resulting bound on H$_2$ thermal pressure, $\log{[T_kn_\text{H2}/(\text{K\,cm}^{-3})]}=3.24$--4.56 (68\% CI, for a half-width of 0.66 dex), is comparable to the COMAP-ERA constraint (68\% CI of $\log{[T_kn_\text{H2}/(\text{K\,cm}^{-3})]}=3.27$--4.50 for a half-width of 0.62 dex), but is arguably more robust as it is obtained without relying on the CO(2--1) auto spectrum.
\section{Discussion}
\label{sec:discussion}
We will discuss first what some of our results mean for the model-dependence of our primary qualitative findings in~\autoref{sec:modeldep}, then how mapping multiple transitions lends greater sensitivity to ISM parameters in~\autoref{sec:whymultiband}.
\subsection{Model-dependence of results}
\label{sec:modeldep}
We have presented two very different models of CO emission in this work. Where Model A models the ISM within a dark matter halo as a homogeneous medium---essentially as if it were a single uniform sphere of molecular hydrogen and carbon monoxide somewhere inside the halo---Model B considers the star-forming phase of the ISM in each halo to be made up of many clouds and effectively builds up the CO luminosity per halo from many copies of the `average' molecular cloud. In part due to their starting points (empirical versus physical), these models ended up making rather different predictions about CO excitation, with the SLED evolving strongly with halo mass for Model A from sub- to super-thermal, but only weakly for Model B and staying sub-thermal for the entire halo mass range considered.

This may be responsible in particular for the differences in the expected constraining power of observations of CO(1--0) alone versus observations of one or both of the higher transitions. In the case of a model like Model A, the CO excitation will depend strongly on the model parameters---for instance, the H$_2$ number density can evolve strongly with halo mass depending on the value of $\alpha$. Information from higher transitions becomes particularly valuable in this case, whereas the ground transition may not reveal as much. However, in the case of Model B, the H$_2$ number density is held constant for each cloud across all clouds in a given halo. As such, we do not expect the CO SLED to evolve as strongly with the other input parameters, and the ground transition is more informative given the physical assumptions implicit in the formulation of the model (as well as more observable than the non-ground transitions due to the sub-thermal excitation).

Yet, the qualitative outcomes of the simulation remain similar between the two models we have presented. In particular, the best outcome always results from the COMAP-ERA and `Triple Deluxe' simulations, significantly outperforming the single-transition survey simulations in parameter recovery even when the raw signal-to-noise ratio is comparable. While we will shortly further discuss the importance of surveying multiple lines, it is important to note before doing so that we were able to quantify the advantage of multi-transition LIM over single-transition LIM for two very different models.

There are further model variations well beyond the scope of this paper, including models predicting CO emission so faint as to be barely detectable if at all even with the full might of the `Triple Deluxe' instrumentation performing in impossibly ideal fashion. However, as we noted above, both models fall in the middle of the pack of models predicting CO emission at late reionisation, so they are certainly not unreasonably optimistic given the state of knowledge about high-redshift molecular gas as examined by a fair sampling of literature within roughly the past decade.

In addition, whether future CO LIM experiments can constrain CO excitation is only one of the three key questions driving this paper. Provided that the CO emission is at a detectable level where a COMAP-ERA or `Triple Deluxe' survey ought to be able to constrain the nature of the emitting galaxies, it is clear that those constraints depend strongly on the ability to probe multiple CO transitions in a way that does not depend extremely strongly or sharply on the CO emission model used for simulation and (mock) inference.
\subsection{The necessity of probing multiple transitions}
\label{sec:whymultiband}
Based on the results of our idealised simulations, we find a twofold need for COMAP to observe CO in multiple rotational transitions. The first is improved signal-to-noise over simply maintaining the 30 GHz instrumentation designed for the COMAP Pathfinder. The fiducial COMAP-EoR proposal~\citep{Breysse22} to add a 15 GHz component to the survey to observe the ground CO(1--0) transition has the particular advantage that it will not be limited by the presence of interloper emission in CO lines from lower redshift, which will be the case for the largest-scale modes of CO(2--1) and CO(3--2). Rejection of interloper emission may be possible through a range of methods proposed in the literature~\cite{Breysse15,Cheng16,LidzTaylor16} but it is best to avoid having interlopers add to the uncertainty in the $z\sim7$ observation in the first place as much as possible. Indeed, in the case of Model B, a noise-free interloper-limited 30 GHz survey would only achieve a signal-to-noise ratio for the CO(2--1) auto spectrum on par with the COMAP-ERA signal-to-noise for the CO(1--0)--CO(2--1) cross spectrum alone.

However, the simulations demonstrate clear differences in the qualitative knowledge obtained from single- versus multi-transition CO LIM detections with similar signal-to-noise. It is true, for instance, that a single-band survey would be sufficient to constrain the overall abundance of CO (in the form of the total CO column density per halo, for instance), particularly if it is possible to reject interloper emission. However, such a survey is insufficient to constrain physical variables beyond this that govern the excitation of CO, such as the H$_2$ number density or thermal pressure, or the kinetic temperature.

Most interesting is the alternate COMAP-ERA concept of adding 45 GHz instrumentation instead of 15 GHz instrumentation. In the case of Model A, where the CO content of the halos becomes sufficiently excited for the CO(2--1) and CO(3--2) fluctuations to be as bright as CO(1--0) fluctuations in aggregate, the combination of Ka- and Q-band data can constrain the kinetic temperature, but is unable to sufficiently constrain H$_2$ number density. In the case of Model B, where the higher transitions are clearly subthermal and thus less observable (by a factor of 2.5 in terms of total signal-to-noise), even the kinetic temperature is not constrained significantly beyond the prior distribution. The story unfolds conversely for hypothetical 15 GHz only surveys, whose constraining power is weaker under Model A and stronger for Model B, although still much weaker than multi-band scenarios in any case.

While differences between the models (as outlined in the previous subsection) clearly result in differences in the constraining power of these scenarios, the common upshot is that CO LIM observations cannot rely on the higher rotational transitions alone to tell the whole story about the high-redshift ISM, certainly not one non-ground transition on its own. Meanwhile, CO(1--0) observations may have some constraining power on their own, but the signal-to-noise required to match constraints from multi-band LIM is outrageously high. In the case of Model B, constraints on $T_k$ and $n_\text{H2}$ from a completely noise-free, sample variance-limited 15 GHz survey---with a total signal-to-noise ratio of 310---essentially has the same constraining power in between a COMAP-ERA 15/30 GHz analysis or a `Triple Deluxe' 15/30/45 GHz analysis, with total signal-to-noise ratios no higher than 70. Despite the equivalent of roughly four to five times higher survey time, the information content of CO(1--0) emission alone is simply not as `dense' as multi-transition data.

Thus, a multi-band approach to future phases of COMAP is a necessity in both practical and scientific terms. Not only will the additional data increase detectability of high-$z$ CO, but also at a given total detection significance across all observables, the constraints obtained on ISM parameters are simply much stronger for multi-band LIM than for single-band LIM.

\section{Conclusions}
\label{sec:conclusions}
Through modelling detection significance and corresponding parameter recovery for a number of signal models and experimental scenarios, we end up with the following answers to the questions we posed in~\href{sec:intro}{the Introduction}:
\begin{itemize}
    \item \emph{What constraining power can future CO line-intensity mapping provide on the environmental variables that govern CO excitation?} In addition to constraining the overall column density of CO hosted in dark matter halos, future phases of COMAP can place lower bounds on the ISM kinetic temperature and upper bounds on the abundance of H$_2$, the chief collisional partner for CO.
    \item \emph{How much does this constraining power depend on the ability of future surveys to observe multiple lines, in particular multiple transitions of CO?} The ability to constrain overall CO abundances per galaxy or proto-cluster (in terms of column density) does not depend strongly on the ability to observe multiple transitions. However, constraints on kinetic temperature, H$_2$ density, or H$_2$ thermal pressure strengthen considerably by measuring auto and cross power spectra across multiple rotational transitions of CO.
    \item \emph{Do qualitative aspects of answers to the above change with our approach to modelling the high-redshift CO signal?} Apart from changes in relative detectability of different transitions, our Model A and Model B simulations provide a consistent qualitative result: observations of a single CO transition alone cannot robustly constrain ISM variables beyond bulk CO abundances. 
\end{itemize}
This work thus demonstrates not only the ability of multi-transition CO LIM as envisioned by COMAP to constrain the characteristics of molecular gas, but also the necessity of the `multi-transition' aspect of the COMAP-EoR, COMAP-ERA, and `Triple Deluxe' survey designs in obtaining such constraints. The endpoint of LIM however, of course, is multi-phase, multi-species LIM, interpreted using physically motivated, self-consistent modelling of a diversity of atomic and molecular gas species (cf., e.g.,~\cite{Sun19}), for a cosmic statistical survey of the high-redshift ISM, CGM, and IGM. Atomic hydrogen lines are key in this endeavour, but so are carbonic lines beyond CO like [C\textsc{\,i}] and [C\textsc{\,ii}], as additional probes of carbon abundance and thus gas metallicity in different ISM phases at high redshift---see, e.g., the analysis of~\cite{Boogaard20} combining CO and [C\textsc{\,i}] line observations with far-infrared continuum data to constrain the properties of photodissociation regions in CO-selected galaxies at $z\sim1$--3.

In devising suitable models of the high-redshift ISM, LIM signal forecasts will need to interface more with both semi-analytic models and hydrodynamical simulations that explicitly simulate processes of cloud formation and dissociation. Of particular value will be the use of Gaussian process emulators (as considered in this work, although hopefully employed with more sophistication) or Gaussian mixture models (cf., e.g.,~\cite{Zhang23,BreysseGMMinprep}) to inform halo models to `paint' onto large-scale cosmological simulations, or dimensionality reduction techniques to gain insights into specific variables that describe the `painting'. The advantage of some of these machine learning models will be the ability to characterise uncertainty in simplifying or extrapolating the fine-grain information, and the ability in principle to propagate these uncertainties all the way through to parameter inferences.

Multi-transition and multi-line LIM scenarios provide compelling science prospects not just for future ground-based observations culminating in arrays of single-dish receivers, but also perhaps for space missions providing multi- or wide-band spectrometer backends, with more reliable access to certain frequency windows and thus certain higher energy transitions of CO and other lines redshifted into sub-cm observing wavelengths. This includes not only concepts that could resolve the LIM signal fluctuations at scales relevant to galaxy formation, but also missions whose primary focus is on the global sky-averaged microwave intensity spectrum (e.g.,~\cite{PIXIE,Delabrouille21}). These observations will contain high information content about the redshift evolution of cosmic line-luminosity densities~\cite{selfcite_inprep} and thus can strongly complement LIM measurements of line-intensity fluctuations, along with measurements of dust and continuum emission in individual objects. Only through a concerted community effort exploiting all such available tracers can we obtain the complete physical picture of star formation at high redshift and the cooling and heating processes that govern it, and the prospects outlined in this work show that multi-transition CO LIM with COMAP would make important contributions to this programme.
\appendix
\section{Effect of select model variations on predictions}
\label{sec:modelvar}
\added{In this section, we consider the effect of select model parameters on the predicted observables via their influence on the halo mass--CO luminosity relation, using Model B as our starting point. We showcase the effect of simple variations on selected parameters in~\autoref{sec:modelvar_basic}, and go on in~\autoref{sec:scatter} to consider how incorporating halo-to-halo scatter in different ways affects our predictions.}

\subsection{Basic variations on Model B}
\label{sec:modelvar_basic}
\begin{figure}
    \centering
    \includegraphics[width=0.5\linewidth]{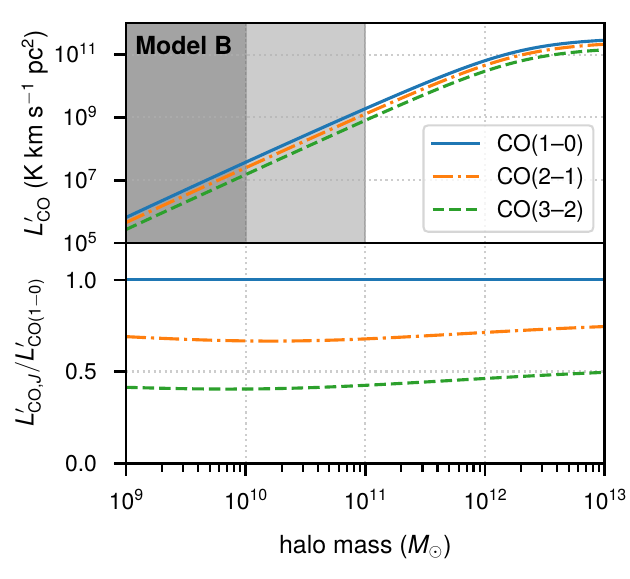}\includegraphics[width=0.46\linewidth]{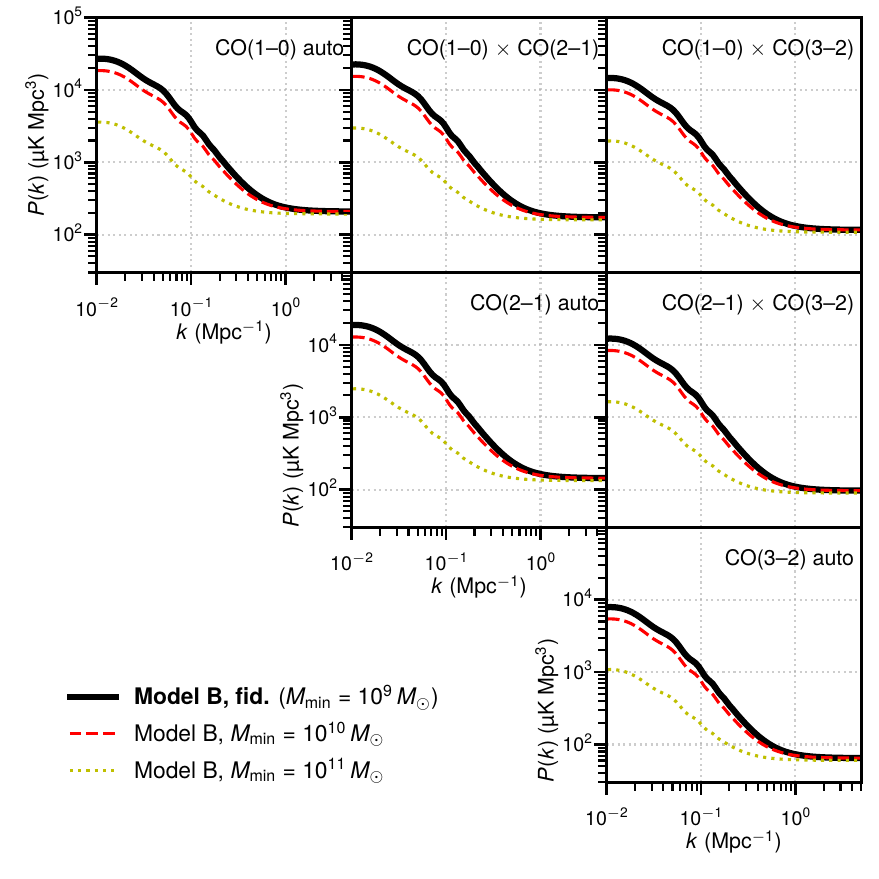}
    
    \includegraphics[width=0.5\linewidth]{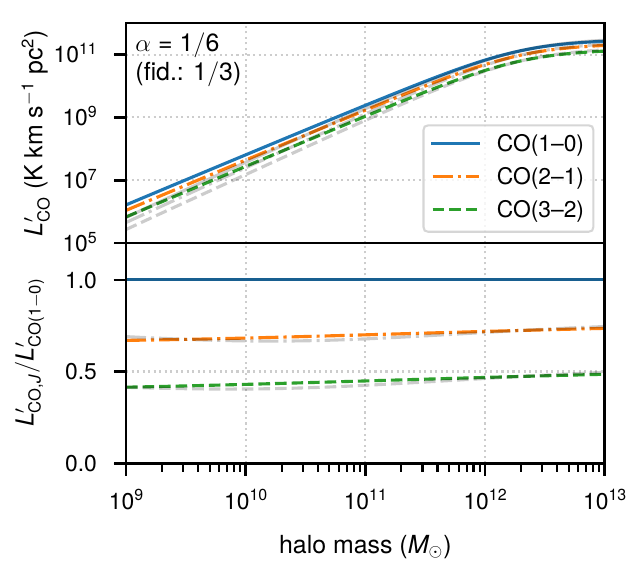}\includegraphics[width=0.46\linewidth]{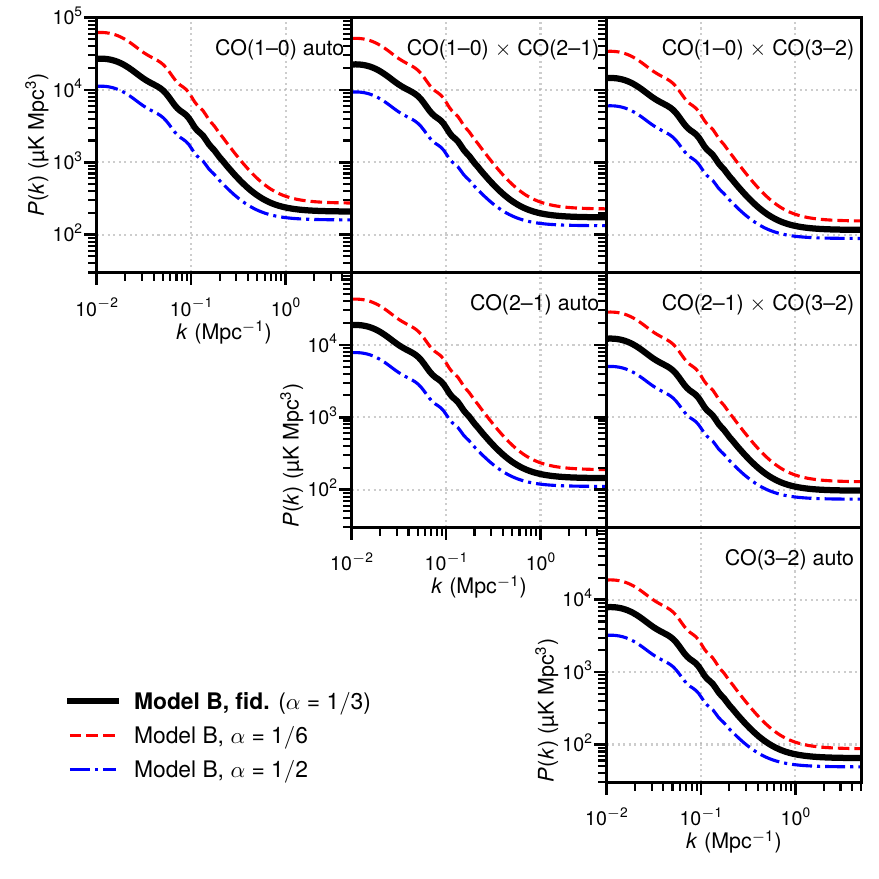}
    \caption{\added{Effect of varying $M_\text{min}$ \emph{(upper panels)} and $\alpha$ \emph{(lower panels)} on the halo mass--line luminosity relations \emph{(left panels)} and the CO auto and cross power spectra \emph{(right panels)}. We indicate the fiducial predictions in fainter grey curves in the left panels, and in thick black curves in the right panels. Note that varying $M_\text{min}$ does not affect the $L'_{\text{CO,}J}(M_h)$ relation above $M_\text{min}$, with the shaded areas in the left panels corresponding to the cutoffs imposed for the power spectra plotted in the right panel.}}
    \label{fig:modelvar_1}
\end{figure}

\added{\autoref{fig:modelvar_1} shows the first kind of variation, varying either the minimum CO emitting halo mass $M_\text{min}$ or the power-law index $\alpha$ controlling the CO column density at low masses. In both cases, the effect is predominantly to increase or decrease the clustering component of all power spectra while leaving the shot noise relatively or completely unaffected.}
\begin{figure}
    \centering
    \includegraphics[width=0.5\linewidth]{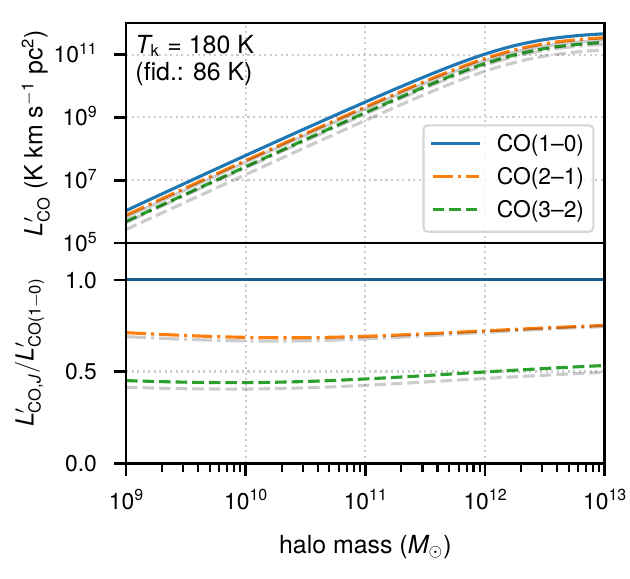}\includegraphics[width=0.46\linewidth]{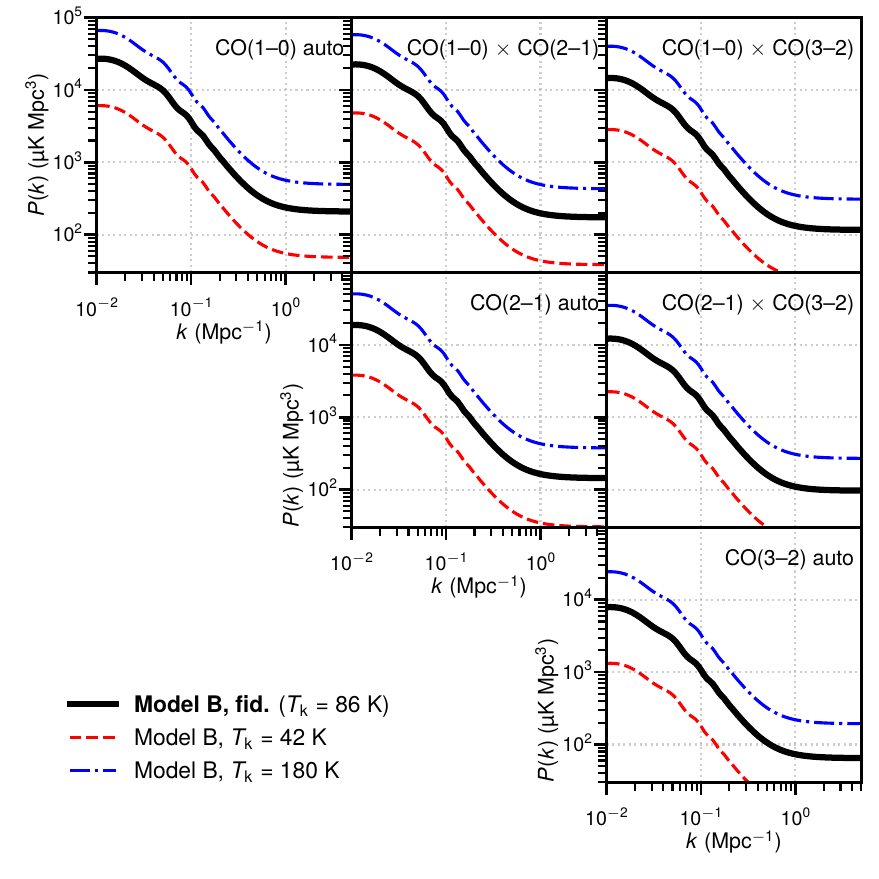}
    
    \includegraphics[width=0.5\linewidth]{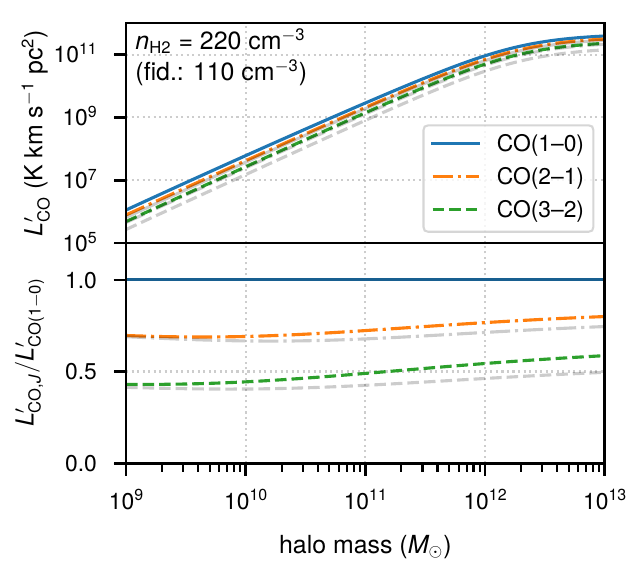}\includegraphics[width=0.46\linewidth]{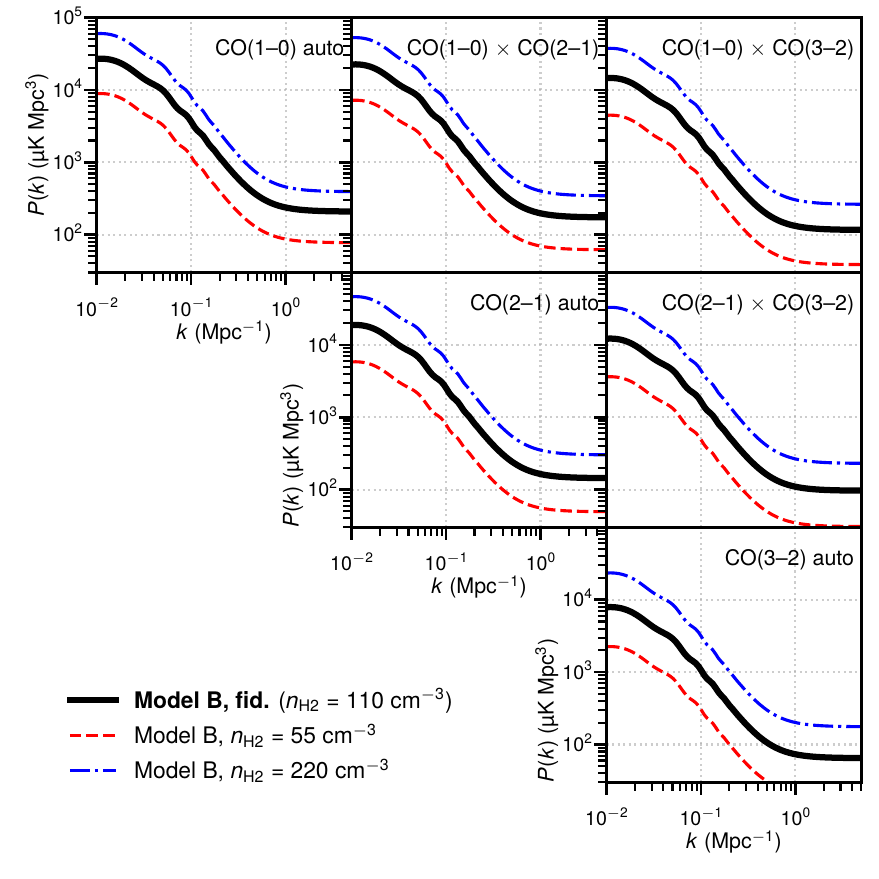}
    
    \includegraphics[width=0.5\linewidth]{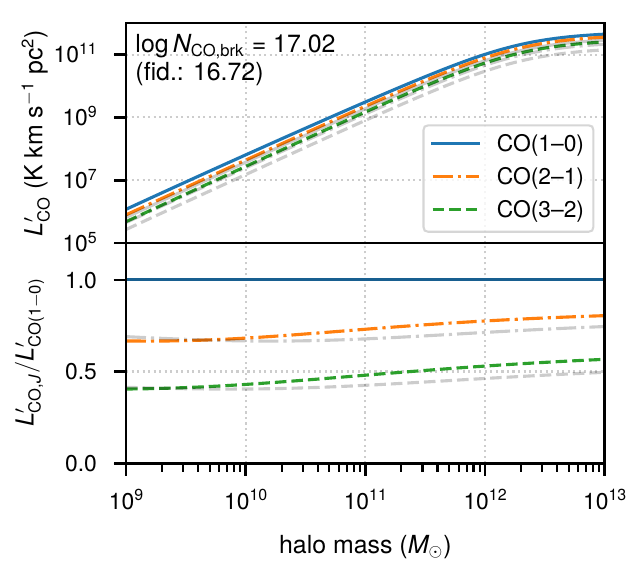}\includegraphics[width=0.46\linewidth]{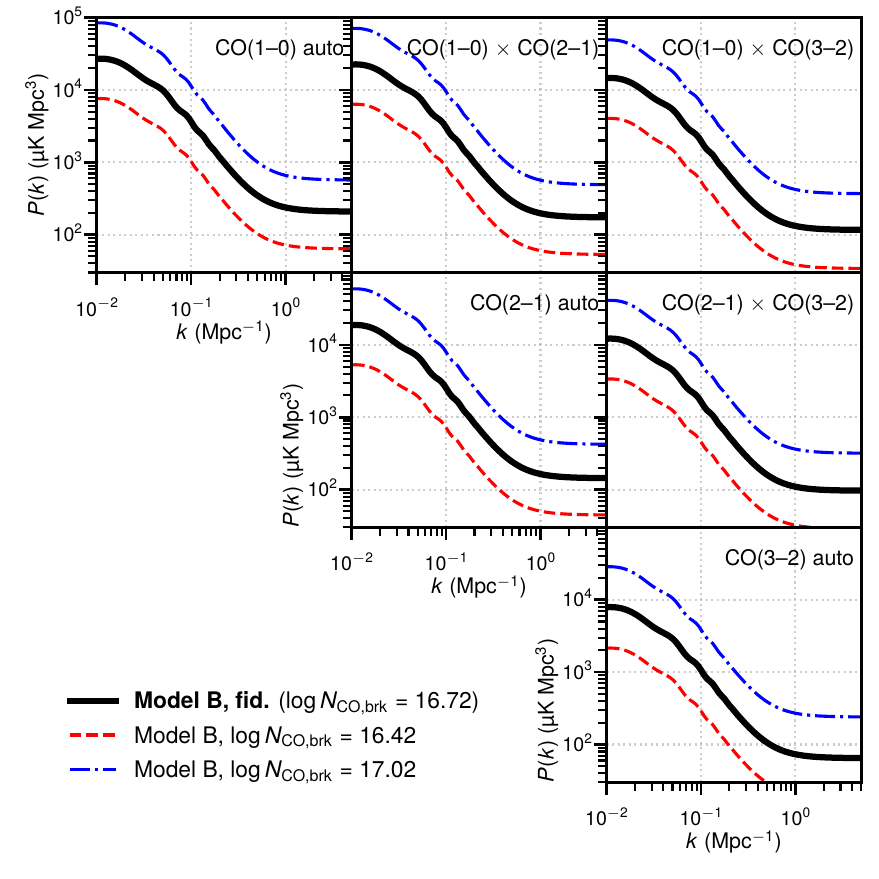}
    \caption{\added{Same as~\autoref{fig:modelvar_1}, but for $T_k$, $n_\text{H2}$, and $N_\text{CO,brk}$ as indicated in each panel.}}
    \label{fig:modelvar_2}
\end{figure}

\added{\autoref{fig:modelvar_2} shows how increasing density and temperature parameters increases the signal. Increasing the kinetic temperature of each cloud, a parameter uniformly applied across halos of all masses, uniformly increases CO excitation and luminosities. The effect is nonlinear however as the observable is the contrast of the line emission against the CMB temperature, rather than the intrinsic line emission. The density parameters $n_\text{H2}$ and $N_\text{CO,brk}$ control the normalisation of the halo mass-dependent CO column density, resulting in non-uniform changes to the SLED. As discussed in the main text, sharp changes in the density parameters are precisely what cause the strong evolution of the SLED with halo mass in Model A.}

\subsection{Effect of log-normal halo-to-halo scatter}
\label{sec:scatter}
\added{If halo-to-halo scatter exists around the mean $L'_J(M_h)$ relation in the form of a log-normal distribution,\footnote{One could instead prescribe cloud-to-cloud stochasticity, but this would require a description of the distribution of the linear sum of many independent log-normal random variates. We leave this to future work.} this elevates the auto shot noise by a multiplicative factor over~\autoref{eq:Pshot_auto}:
\begin{equation}
    P_{s,J} = \exp{\left(\sigma_{\text{LN},J}^2\ln^2{10}\right)}\frac{(1+z)^4}{H^2(z)}\int dM_h\,\frac{dn}{dM_h}\,L^{\prime2}_J(M_h).\label{eq:Pshot_auto_sc}
\end{equation}
Here $\sigma_{\text{LN},J}$ describes the width of the log-normal distribution in units of dex.

A similar prefactor can enter cross shot noise calculations, but only under the assumption of nonzero correlations between the line luminosities $L'_J(M_h,z)$ and $L'_{J'}(M_h,z)$ from halo to halo. Per~\cite{Yang22}, assuming a cross-correlation coefficient $\rho_{J,J'}$ between the CO($J\to J-1$) and CO($J'\to J'-1$) transitions, this introduces a multiplicative factor in front of~\autoref{eq:Pshot_cross}:
\begin{align}
    P_{s,J\times J'} &= \left\{1+\rho_{J,J'}\left[\exp{(\sigma_{\text{LN},J}^2\ln^2{10})}-1\right]^{1/2}\left[\exp{(\sigma_{\text{LN},J'}^2\ln^2{10})}-1\right]^{1/2}\right\}\nonumber\\&\hspace{2.86cm}\times\frac{(1+z)^4}{H^2(z)}\int dM_h\,\frac{dn}{dM_h}\,L'_J(M_h)L'_{J'}(M_h).
\end{align}
This readily reduces to~\autoref{eq:Pshot_auto_sc} when $J=J'$ (since $\rho_{J,J}=1$).
}

\begin{figure}
    \centering
    \includegraphics[width=0.86\linewidth]{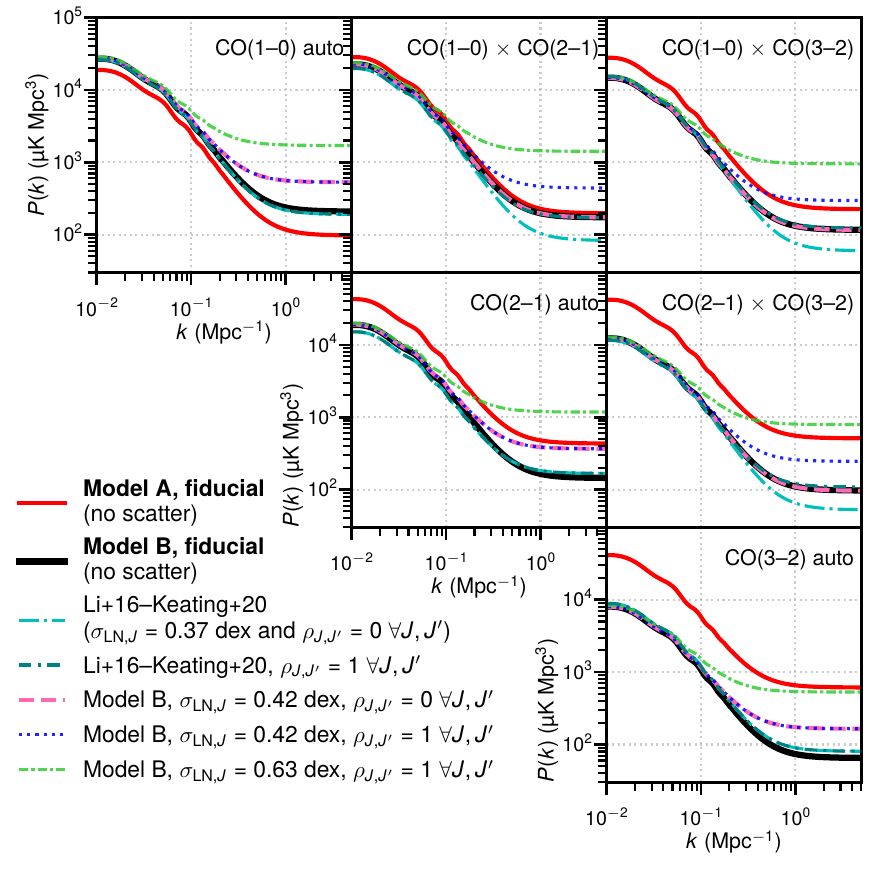}
    \caption{Same as~\autoref{fig:models}, but with the addition of many variations that tweak prescriptions around halo-to-halo scatter affecting the shot noise components of the auto and cross power spectra, with the inter-line cross-correlation coefficient $\rho_{J,J'}$ exclusively affecting cross power spectra.}
    \label{fig:modelsc}
\end{figure}

\added{We show in~\autoref{fig:modelsc} the effect of introducing scatter to Model B, comparing the resulting predictions to each other as well as to the predictions of the `Li+16--Keating+20' model~\cite{Li16,mmIME-ACA} considered in~\cite{Breysse22} and discussed in this work in~\autoref{sec:snr}. This `Li+16--Keating+20' model (which we denote L16K20 below for brevity) by default assumes $\rho_{J,J'}=0$ between $J=1$ and $J'=2$, and we follow this choice in the main text where we assume $\rho_{J,J'}=0$ for all $J\neq J'$. However, in~\autoref{fig:modelsc} we show the effect of assuming non-zero inter-line correlations by showing a variation of the model setting $\rho_{J,J'}=1$ for all $J\neq J'$.}

\added{This in turn shows that the cross power spectra for Model B only fall off more slowly at high $k$ than the L16K20 model because of assuming $\rho_{J,J'}=0$ for all $J\neq J'$, which suppresses the cross shot noise. In fact, the second moment of CO line luminosities is intrinsically higher throughout for Model B than for L16K20. The lack of scatter in Model B compensates for this in the auto shot noise, but not in the cross shot noise unless we set $\rho_{J,J'}=1$ in L16K20 for all $J\neq J'$. Differences between Model A and Model B, on the other hand, are entirely due to differences in second moments of CO line luminosities. The mass-dependent CO SLED of Model A results in different ratios of shot noise to clustering amplitude across the CO line ladder, whereas the power spectrum shape remains the same across the ladder for Model B.}

\added{We also show in~\autoref{fig:modelsc} that by introducing halo-to-halo scatter with or without inter-line correlations to Model B, we control the amplitude of the auto and cross (or just auto) shot noise and thus the shape of the power spectra. Within the range of $k\in(0.03,0.5)\,$Mpc$^{-1}$ considered for the primary results in the main text, the effect is mostly confined to bins of higher $k$ and even there at most an order of unity for a value of $\sigma_{\text{LN},J}=0.42$ dex, on a level similar to what L16K20 assumes, though larger for a more extreme value of 0.63 dex. We omit consideration of stochasticity from the main text here, but it will be more relevant for future work that specifically examines statistics at small scales and/or beyond $P(k)$.}

\section{Interloper model: details and comparison to observational constraints}
\label{sec:interloper}
For all interloper line emission, the basic form of the luminosity function assumed is the combination of a quasi-exponential low-luminosity cutoff and a Schechter function, itself the combination of a power law and an exponential high-luminosity cutoff:
\begin{equation}
    \phi(L)\equiv \frac{dn}{dL} = \frac{\phi^*}{L^*}\left(\frac{L}{L^*}\right)^\alpha\exp{\left(-\frac{L}{L^*}-\frac{L_\text{min}}{L}\right)}.\end{equation}
The original parameter values tuned to match the CO(1--0) luminosity function of~\cite{Li16} are $\phi^*/L^*=8.7\times10^{-11}\,$Mpc$^{-3}\,L_\odot^{-1}$, $L^*=2.1\times10^6\,L_\odot$, $\alpha=-1.87$, $L_\text{min}=500\,L_\odot$. However, here we make a subtle change, which is to recast the luminosity function in terms of observer units of velocity- and area-integrated brightness temperature:
\begin{equation}
    \phi(L')\equiv \frac{dn}{dL'} = \frac{\phi^*}{L^{\prime*}}\left(\frac{L'}{L^{\prime*}}\right)^\alpha\exp{\left(-\frac{L'}{L^{\prime*}}-\frac{L'_\text{min}}{L'}\right)}.\end{equation}
In this space, we still have $\alpha=-1.87$, but $\phi^*/L^{\prime*}=4.3\times10^{-15}\,$Mpc$^{-3}\,($K km s$^{-1}$ pc$^2)^{-1}$, $L^{\prime*}=4.2\times10^{10}$ K km s$^{-1}$ pc$^2$, and $L'_\text{min}=10^7$ K km s$^{-1}$ pc$^2$. We then apply these same parameters as necessary to the $z\sim1.6$ CO(1--0) and $z\sim4$ CO(2--1) interloper components that affect the 45 GHz observation.

\begin{figure}
    \centering
    \includegraphics[width=\linewidth]{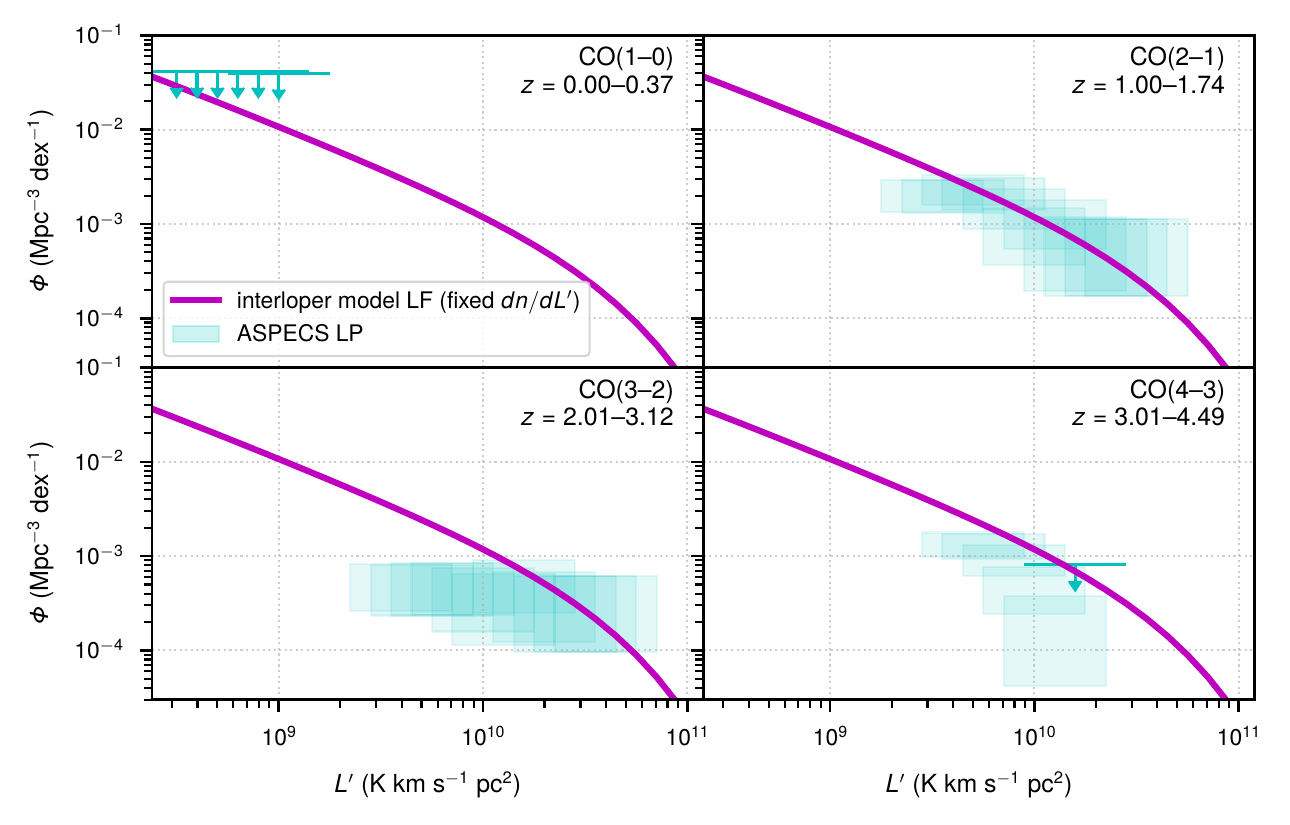}
    \caption{The luminosity function assumed for all interloper line emission components, compared against all constraints on CO luminosity functions derived from 3\,mm ASPECS LP data~\cite{Decarli20}. For completeness we also show ASPECS LP upper limits on the CO(1--0) luminosity function at $z\sim0$, although not strictly relevant to this particular work.}
    \label{fig:schcut}
\end{figure}

\autoref{fig:schcut} provides a comparison of this luminosity function (now as $\Phi(L')\equiv dn/d(\log{L'})$ instead of the original $\phi=dn/dL'$) against constraints derived from 3\,mm observations of the ALMA SPECtroscopic Survey Large Programme (ASPECS LP) in the Hubble Ultra-Deep Field~\cite{Decarli20}. We see that the same luminosity function is broadly consistent with all constraints on all observed transitions, with particularly good agreement with the observed CO(2--1) luminosity function at $z\sim1.0$--1.7. Deviations exist but are within an order of magnitude. Note also that each one of our interloper transitions is one rung below the transition observed at the same redshift in the 3\,mm ASPECS LP data; even somewhat subthermal excitation could mean sizeable swings in luminosities further up the ladder of rotational transitions.
\section{Monte Carlo posterior distributions in original parameter spaces}
\label{sec:mcmc_full}
\begin{figure}
    \centering
    \includegraphics[width=0.6986\linewidth]{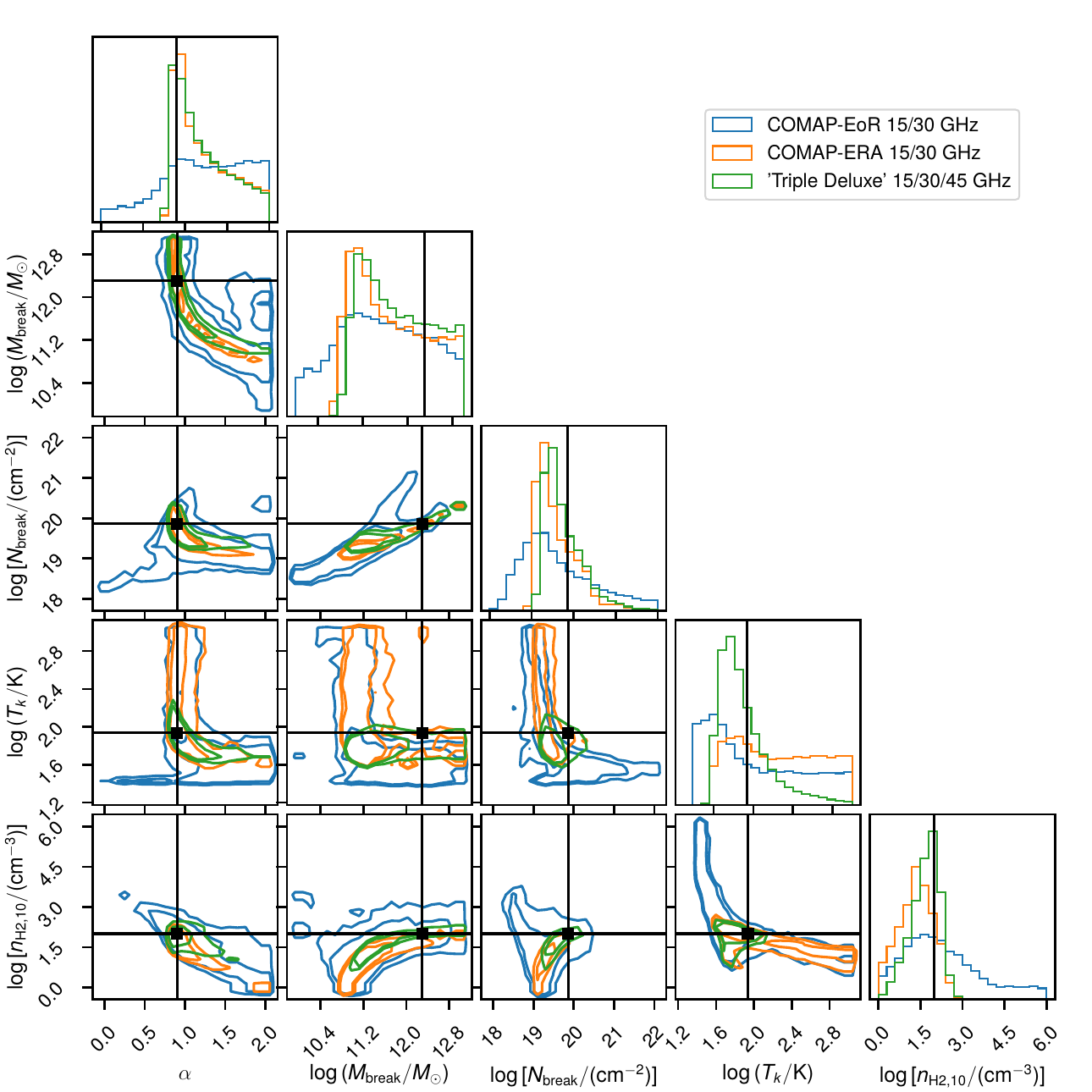}
    \includegraphics[width=0.6986\linewidth]{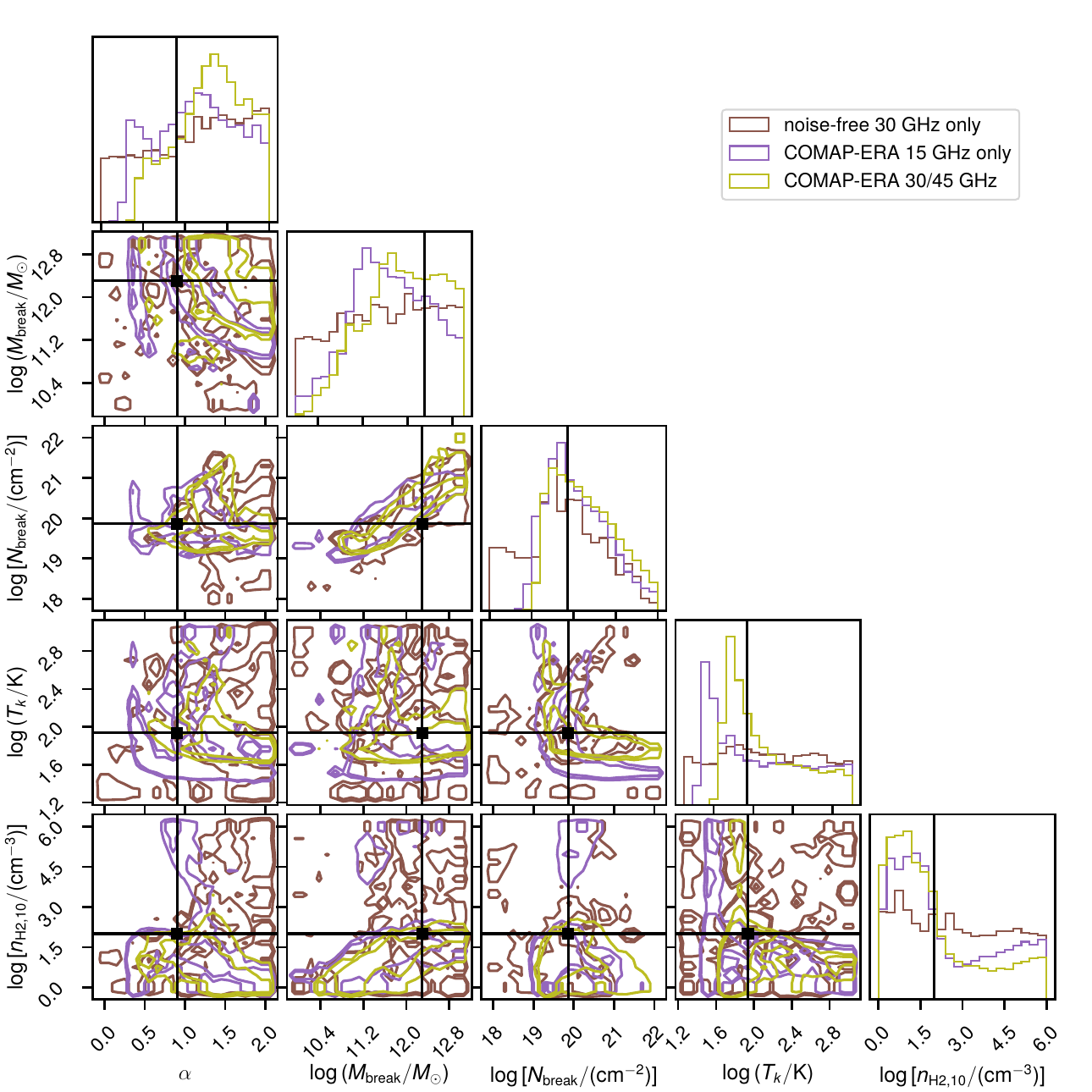}
    \caption{Full MCMC posterior distributions for exercises carried out with Model A.}
    \label{fig:modelA_corner}
\end{figure}

\begin{figure}
    \centering
    \includegraphics[width=0.6986\linewidth]{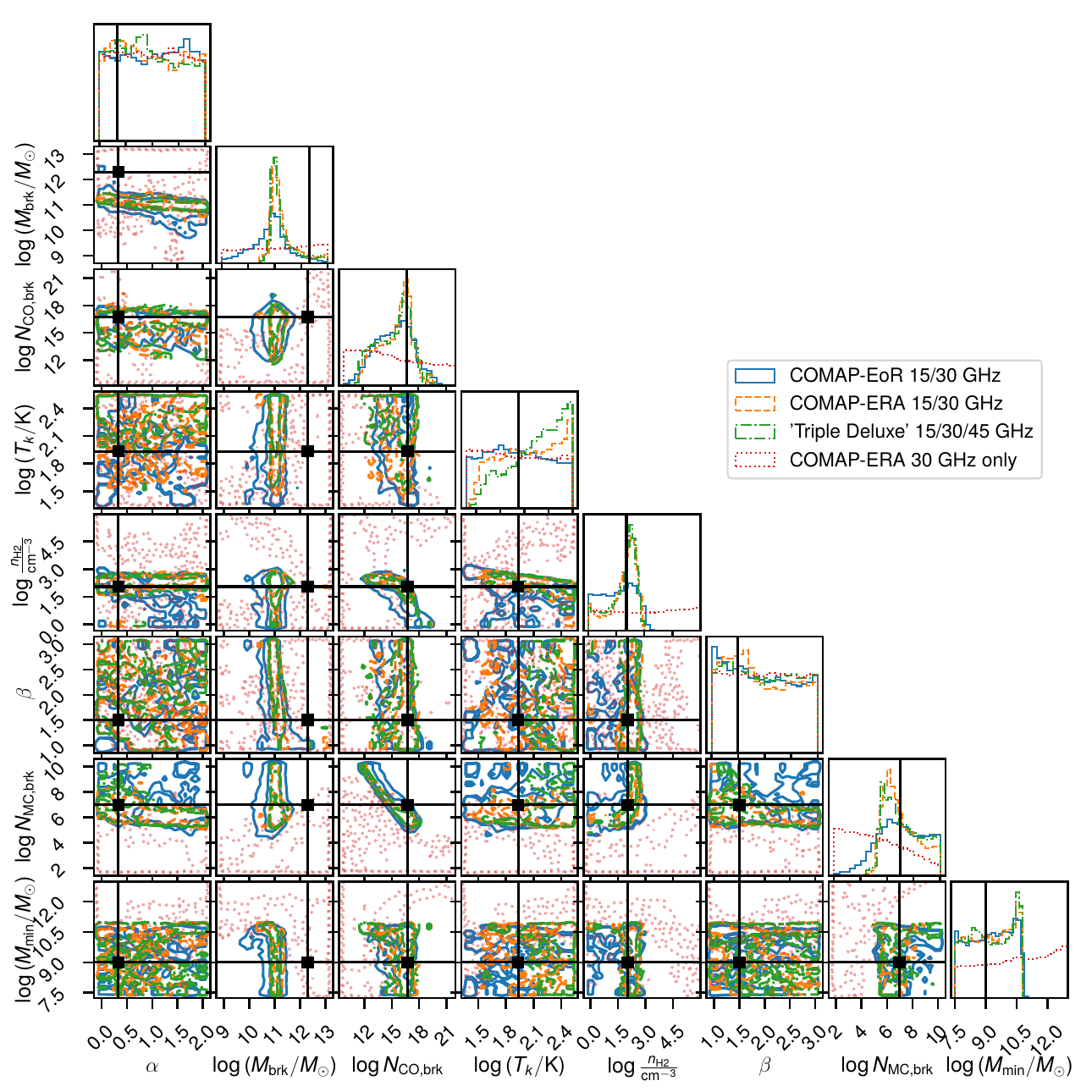}
    \includegraphics[width=0.6986\linewidth]{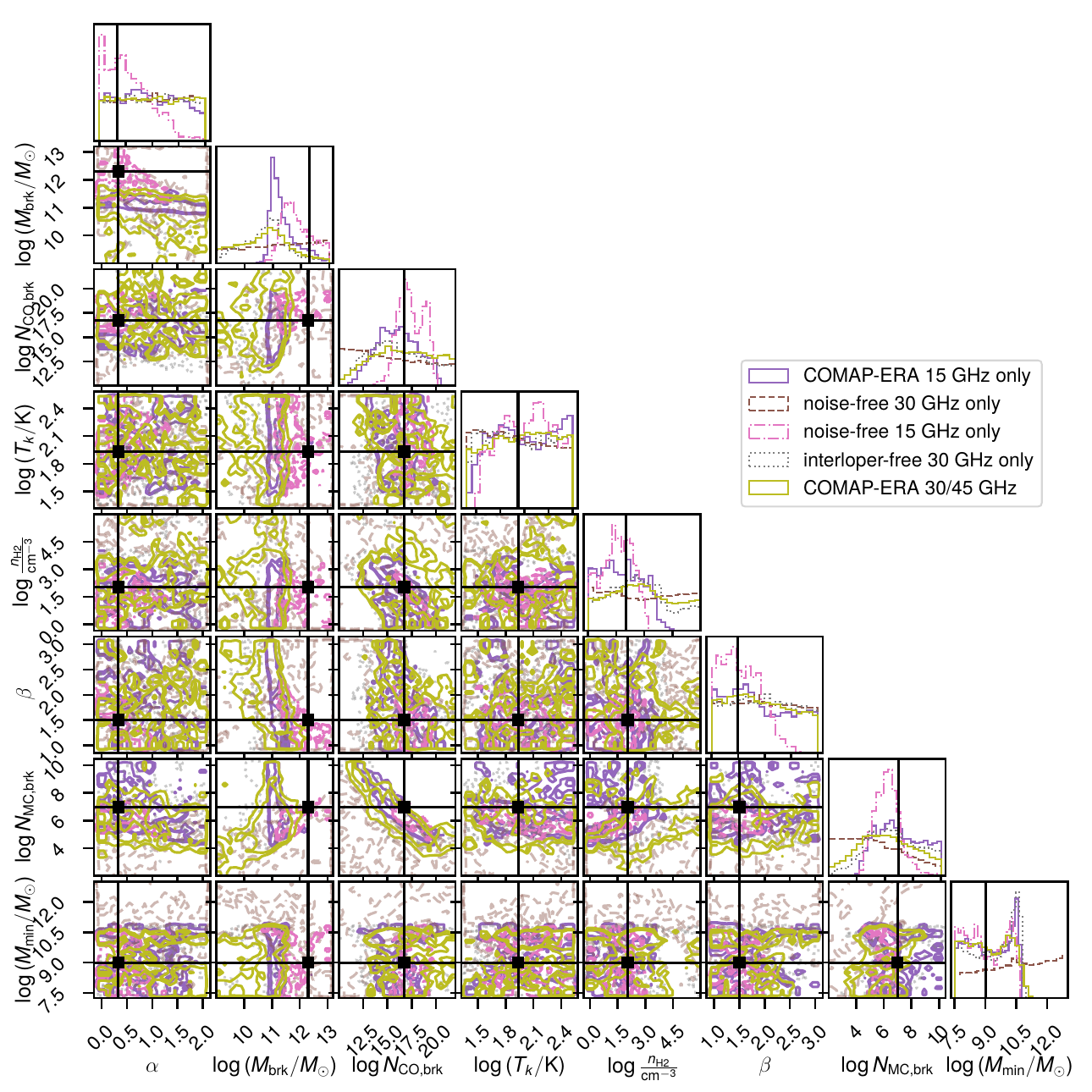}
    \caption{Full MCMC posterior distributions for exercises carried out with Model B.}
    \label{fig:modelB_corner}
\end{figure}

\autoref{fig:modelA_corner} and \autoref{fig:modelB_corner} show the full MCMC `corner plots' for all exercises in the main text, for Models A and B respectively. For Model A, the difference between the multi- and single-band MCMC exercises are evident in the bounded, mostly well-defined contours for the multi-band scenarios and the foamy flat distributions for the single-band (and Ka-/Q-band) scenarios. For Model B, we find that even the multi-band scenarios cannot provide significant constraints on certain parameters, like the power-law indices $\alpha$ and $\beta$ for the CO column density and molecular cloud count, but otherwise we encounter a similar narrative. The exception, oddly enough, is that the noise-free CO(1--0)-only scenario can successfully bound both power-law indices for Model B from above, owing to its sensitivity to at least the overall abundance of CO.

In the case of Model B, we find that the relevant mass scales are largely misidentified, with the minimum mass for CO emission unconstrained, and $M_\text{brk}$ identified as $\sim10^{11}\,M_\odot$ instead of the fiducial value, which is around 20 times that. We ascribe this to the fact that the observation senses not all halos above a certain minimum mass or all halos below a certain maximum mass, but rather whatever halos dominate the signal. Indeed, despite misidentification of the break mass, we find in~\autoref{fig:modelB} a successful recovery of the product of cloud count and CO column density at these halo masses. Still, this suggests that in practice, when the break in a double power law is unlikely to affect the signal---which~\autoref{fig:dTdM} suggests is the case---fitting a single power law may be preferable so as to reduce complexity without sacrificing information recovery.

We also note that analysing the CO(1--0) data by itself tends to result in a downward bias of $n_\text{H2}$ from the `true' value, whereas analysing one or both of the higher transitions results in a very weak upward bias from the `true' value (although some of this may be due to our emulator limits as discussed in~\autoref{sec:emulator}). This once again highlights the need for multi-transition CO LIM; adding further integration time and further noise may well reduce the remaining (albeit much reduced) bias in the COMAP `Triple Deluxe' posterior.

\acknowledgments

Research in Canada is supported by NSERC and CIFAR. Parts of these calculations were performed on the Sunnyvale cluster and other high-performance computing nodes at CITA; the author thanks John Dubinski and MJ Huang for tireless maintenance of these resources.

Thanks to members of the COMAP Collaboration, including Kieran Cleary, Patrick Breysse, Hamsa Padmanabhan, and Do\v{g}a Tolgay, for valuable suggestions and detailed comments that led to the initiation and completion of this work.

DTC is supported by a CITA/Dunlap Institute postdoctoral fellowship. The Dunlap Institute is funded through an endowment established by the David Dunlap family and the University of Toronto. The University of Toronto operates on the traditional land of the Huron-Wendat, the Seneca, and most recently, the Mississaugas of the Credit River; DTC is grateful to have the opportunity to work on this land. DTC also acknowledges support through the Vincent and Beatrice Tremaine Postdoctoral Fellowship at CITA.

This work made use of \texttt{astropy},\footnote{http://www.astropy.org} a community-developed core Python package and an ecosystem of tools and resources for astronomy \citep{astropy:2013, astropy:2018, astropy:2022}. This research also made use of NASA's Astrophysics Data System Bibliographic Services.



\bibliographystyle{JHEP}
\bibliography{bibliold,biblio}

\end{document}